\documentclass[12pt,fleqn,psfig]{article}
\usepackage{amsmath, amssymb, graphicx}
\usepackage[hang,scriptsize,bf]{caption}
\usepackage{cite}
\usepackage{hyperref}
\usepackage{geometry}
\usepackage{color}
\usepackage[latin1]{inputenc}
\usepackage{mathrsfs}
\usepackage{amsfonts}
\usepackage{subcaption}

\usepackage{fancyhdr}
\begin{document}

\begin{center} \noindent 
	{\bf  Lapse singularities, caustics and entanglement}\end{center}


\bigskip\bigskip\bigskip

\bigskip\bigskip\bigskip

\begin{center}
Zachary Guralnik
\footnote{Zachary.S.Guralnik@leidos.com}\\
\it \small Leidos, Inc. 11951 Freedom Dr, Reston, VA, 20190
\end{center}
\bigskip

\bigskip\bigskip\bigskip\bigskip

\renewcommand{\thefootnote}{\arabic{footnote}}

\centerline{ \small Abstract}
\bigskip

\small 

We study diffraction catastrophes of wave functions in diffeomorphism invariant quantum theories, for which $\hat H\Psi=0$.  These wave functions can be represented in terms of integrations over cycles in a complexified lapse variable $N$.   The integrand $\exp(i{\mathbb S}(N))$ may have multiple essential singularities  at finite values of $N$ and at infinity.  A basis set for Greens functions and solutions of the wave equation is represented by Lefschetz thimbles connecting these singularities.   The finite $N$ singularities are shown to be directly related to $A_{n\ge 3}$ caustics. We give an example similar to a minisuperspace cosmological model constructed by Halliwell and Myers, to which we add a scalar field. We show that caustics with codimension $d\ge 2$ exhibit strong entanglement with respect to partitions of their unfolding degrees of freedom. If an unfolding direction corresponds to a physical clock in a solution of the Wheeler-DeWitt equation, the caustic bears some resemblance to a quantum measurement. The R\'enyi entanglement entropy ${\cal R}_n$  is expressed in terms of integrals over $2n$ lapse variables $N_i$. Writing the integrand as $\exp(i\Gamma)$, we find that the  finite $N$ essential singularities of $\exp(i{\mathbb S})$ are replaced with non-essential singularities of $\exp(i\Gamma)$ at cyclically related $N_i = N_j$ ,  which the Lefschetz thimbles evade.  The relative homology classes to which the integration cycles belong  are higher dimensional variants of links.

\newpage
\section{Introduction}
The vanishing Hamiltonian constraint in the ADM formulation of quantum gravity \cite{ADM} is imposed by a Lagrange multiplier $N(x^\mu)$ known as the lapse function. In the minisuperspace approximation, the path integral representation of the wavefunction has the form,  
\begin{align}\label{forml}
\Psi(\bf q)=\int {\cal D}{\bf P}{\cal D}{\bf Q} {\cal D}{N}  e^{    
         i \int d\tau \left(  {\bf P}\cdot{\dot{\bf Q}} - N H({\bf P},{\bf Q}) \right)
	 } \, , 
\end{align}
with ${\bf Q}(1)=\bf q$ and $\tau$ defined on the interval $[0,1]$.  The variables $\bf Q(\tau)$ represent spatial zero modes of various fields in the full gravitational theory.  
In the gauge $dN/d\tau=0$, 
integrating over ${\bf P}$ and ${\bf Q}$ leaves a one dimensional integral,
\begin{align} 
\Psi = \int d N e^{i
		   {\mathbb S}(N) 
} \, ,
\end{align}
where ${\mathbb S}$ will be referred to as the lapse action.
The physical implications of the analytic structure of the lapse action are the subject of this article.
In particular, the finite $N$ singularities of $\mathbb S$ will be shown to be related to the ``non-smooth''caustics, which are those with codimension greater than one. The behavior of $\Psi$ in the neighborhood of caustics is described by a set of standard forms known as diffraction catastrophes.  In the case of a Euclidean path integral, an analytic continuation of these standard forms arises, however we shall mostly consider the Lorentzian case.

The set of solutions to $\hat H\Psi=0$ is determined by the essential singularities of $\exp( i {\mathbb S}(N)   )$, which serve as boundaries of integration in the complex $N$ plane.
Equivalence classes of convergent integration contours in the complex $N$ plane have representative steepest descent paths, or Lefschetz thimbles, on which the phase ${\rm Re} ({\mathbb S})$ is constant.    Complexified path integration and Lefschetz thimbles have been discussed in numerous contexts, with some representative examples found in \cite{Gibbons,HalliwellHartle,GG1,GG2,Pehlevan,Guralnik,WIT,Ferrante,Dunne,Behtash,Cristoforetti,TurokLorentzian, TurokRadio}. 
The lapse action is non-polynomial 
and may have multiple finite poles and other singularities such that $\exp(i{\mathbb S}(N))$ has essential singularities at finite $N$.   The path integral yields either solutions of the Hamiltonian constraint or Greens function of the Hamiltonian, depending on which sums over Lefschetz thimbles with finite $N$ boundaries are included in the integration. 

Finite $N$ essential singularities of $\exp(i{\mathbb S})$ originate in a collapse of the classical solution set obtained by considering variations with respect to $\bf Q$ and $\bf P$ but not $N$.  At particular values $N=N_p$, solutions exist only for a restricted set of boundary conditions.  In the simplest examples, $N_p$ 
are poles of ${\mathbb S}$. 
The residues are a function of the boundaries, $f( {\bf Q}(0),{\bf Q}(1))$, which vanish as the admissible boundary condition are approached.  The locus of vanishing residue is a source if a single Lefschetz thimble ends at the pole, but still has physical implications when two thimbles terminate with opposite orientation at the pole, giving the sum of a source and a sink.  In this case $r=0$ defines a curve to which $A_{n\ge 3}$ caustics are bound, which we shall refer to as a ghost source. Explicit examples will be given for the $A_3$ or ``cusp'' caustic.
The focal property of these caustics is related to the collapse of admissible boundary conditions. 
The relation between the finite $N$ essential singularities of $\exp(i\mathbb S)$ and caustics will be described in section \ref{polesandcaustics}. This relation was originally conjectured to exist and discussed in a different context in \cite{G1,G2}, wherein ${\mathbb S}$ was referred to as the ``einbein action''.

In the neighborhood of a caustic, the wavefunction is described by a diffraction catastrophe having an exponential integral representation \cite{BerryThom},
\begin{align}
\Psi= \int d\boldsymbol{\lambda} \exp( i {\cal P}(\boldsymbol\lambda,\boldsymbol\zeta) )
\end{align}
where ${\cal P}$, known as the potential function, is a  polynomial in $\boldsymbol\lambda$ with coefficients $\boldsymbol\zeta$. The number of variables $\boldsymbol \lambda$ and $\boldsymbol\zeta$ are the corank and codimension of the caustic respectively.   
A classification of the stable caustics according to these polynomials and an associated coxeter group  is due to Arnold \cite{Arnold}, following upon the work of Thom \cite{Thom}. 
Table \ref{AD_catast} lists the stable caustics of codimension $d\le 4$.
Critical points, or $\boldsymbol \lambda$ at which $\frac{\partial{\cal P}}{{\partial \lambda_i}}=0$, coalesce at $\boldsymbol\zeta =0$ in which case the matrix $\frac{\partial^2{\cal P}}{\partial\lambda_i \partial\lambda_j}$ is singular.  The parameters $\zeta_{1\cdots d}$ are the unfolding directions of the caustic. The higher order caustics unfold into lower order caustics,  meaning that the latter can be found at non-zero $\boldsymbol\zeta$.   For instance the codimension two $A_3$ caustic, at which three critical points coalesce, is the common boundary of two codimension one $A_2$ caustics, along which pairs of critical points coalesce.

\begin{table}
	\begin{center}
		\begin{tabular}{|l|l|l|}
			\hline
			$A_2$ & Smooth or Fold Caustic & $\frac{1}{3}\lambda^3 + \zeta_1\lambda$ \\ 
			\hline
			$A_3$ & Cusp Caustic & $\frac{1}{4}\lambda^4 + \frac{1}{2}\zeta_2 \lambda^2 + \zeta_1\lambda$ \\
			\hline
			$A_4$ & Swallowtail Caustic & $\frac{1}{5}\lambda^5 + \frac{1}{3}\zeta_3\lambda^3 + \frac{1}{2}\zeta_2\lambda^2 + \zeta_1\lambda$ \\
			\hline
			$A_5$ & Butterfly Caustic & $\pm\frac{1}{6}\lambda^6 + \frac{1}{4}\zeta_4\lambda^4 + \frac{1}{3}\zeta_3\lambda^3 + \frac{1}{2}\zeta_2\lambda^2 + \zeta_1\lambda$ \\
			\hline
			$D^+_4$ & Hyperbolic Umbilic & $\lambda_1^3 + \lambda_2^3 - \zeta_3\lambda_1\lambda_2-\zeta_2\lambda_2-\zeta_1\lambda_1$ \\ 
			\hline
			$D_4^-$ & Elliptic Umbilic & $\lambda_1^3 - 3\lambda_1\lambda_2^2-\zeta_3(\lambda_1^2+\lambda_2^2)-\zeta_2\lambda_2-\zeta_1\lambda_1$ \\
			\hline
			$D_5$ & Parabolic Umbilic & $\lambda_1^4+\lambda_1\lambda_2^2+\zeta_4\lambda_2^2+\zeta_3\lambda_1^2 + \zeta_2\lambda_2 + \zeta_1\lambda_1$ \\
			\hline	
		\end{tabular}\caption{A list of stable caustics of codimension $d\le 4$ including their potential functions and associated coxeter groups.}\label{AD_catast}
	\end{center}
\end{table}

Unlike the potential functions, which are unique to a particular caustic, the lapse action may capture a network of caustics of different types.  The lapse action  ${\mathbb S}(N)$ is related to a potential function ${\cal P}(\lambda)$  by a locally defined map $\lambda(N)$, valid only for a region of $N$ in the neighborhood of a particular caustic.  The map $\lambda(N)$ turns out to be singular for ${\boldsymbol \zeta}$ on two curves which intersect at the caustic, one of which is the ghost source.    
  
The $D_n$ and $E_n$ catastrophes have corank $>1$. Thus is not clear how they could exist in a diffeomorphism invariant theory expressed as a path integral over a single lapse variable, although perhaps they can be found in a Wheeler-DeWitt wave function expressed as an integral over both lapse and shift variables. The latter are generally set to zero in minisuperspace, but could be of interest in this context.      


Halliwell and Myers have given an example of a Wheeler-DeWitt wave function in a (2+1) dimensional cosmology \cite{HalliwellMyers} reviewed in section \ref{examples}, for which the lapse action ${\mathbb S}(N)$ has an infinite number of finite $N$ simple poles and a pole at infinity. Each critical point lies on a Lefschetz thimble with boundaries at the poles,  
corresponding to a geometry with various numbers of three spheres connected by wormholes. 
Despite the presence of the finite $N$ poles of $\mathbb S$,  there are no caustics with codimension greater than one in the Halliwell-Myers model since  the wave function has only one degree of freedom, the scale factor $a$. 
However there is a precursor of $A_3$ (cusp) caustics which can be seen in the collapse of the classical solution set responsible for the existence of the poles. Cusp caustics are present upon adding a scalar field, or clock variable, $T$  to the model. Due to its derivation from a Euclidean signature path integral, the wave function $\Psi(a,T)$ does not exhibit  the usual $A_3$ diffraction catastrophe, i.e. the Pearcey function, but rather its analytic continuation. The cusps are bound to ghost sources at locations predicted by arguments given in section \ref{polesandcaustics}. 



Caustics are ordinarily thought of as regions where the amplitude of the wave function becomes very large, but have an additional distinguishing feature for codimension $d\ge 2$.   If the path integral is Lorentzian, such that one considers the usual diffraction catastrophe as opposed to an analytic continuation,  the caustic is accompanied by a skeleton of wave function dislocations, or vortices, centered about points where the wave function vanishes \cite{BerryNye1}.  
Entanglement between unfolding degrees of freedom implies non-zero $\Phi_{12} \equiv \partial_1\partial_2\ln\Psi(\zeta_1,\zeta_2,\cdots)$  which diverges at the dislocations, around which 
$\oint d(\ln \Psi) = 2\pi n$. The R\'enyi entanglement entropy can be written as a non-local functional of $\Psi$, described in section \ref{disloc}, with large contributions due to a collusion between regions of large field and dislocations.  The dislocations form networks of strings for caustics of codimension $d=3$.  In this case one can define tripartite entanglement entropy, with substantial contributions due to the presence of both string junctions and regions of large field. 

Since the Wheeler-DeWitt wave function is static, satisfying $\hat H\Psi=0$, time evolution is only defined in terms of entanglement between a physical clock and the other degrees of freedom\cite{PageWooters,Wooters}. 
A codimension $d\ge 2$ caustic for which one of the unfolding directions is a physical clock has properties reminiscent of a quantum measurement, discussed in section \ref{clocstics}. The ``instant of measurement'' maps to a singularity of lapse action. 
The lapse integral representation of the R\'enyi entanglement entropy  takes the form
\begin{align}\label{Rii}
{\cal R}_n = \frac{1}{1-n}{\rm ln} \int dN_1 \cdots dN_{2n} \exp(i\Gamma(N_1 \cdots  N_{2 n}))\, ,
\end{align}
where, as shown in section \ref{ThimbEntangle}, the `R\'enyi' action $\Gamma(N_1 \cdots N_{2n})$ inherits some but not all of its singularity structure from the lapse action $\mathbb S(N)$.  
We shall focus on simple examples in which there is a single ghost source determining a natural partition of degrees of freedom along tangential and transverse directions.  The essential singularities of $\exp(i\Gamma)$  at infinity are trivially determined by those of $\exp(i{\mathbb S})$.  However the finite $N$ essential singularities of $\exp(i{\mathbb S})$ 
are replaced by non-essential singularities of $\exp(i\Gamma)$ at $N_i=N_j$, for cyclically related $i$ and $j$. 
Lefschetz thimbles associated with $\exp(i\Gamma)$ terminate at essential singularities at infinity, but evade the non-essential singularities at $N_i=N_j$.
The integration cycles belong to relative homology classes which are higher dimensional versions of  links. A relation between knots, links and entanglement was also described in \cite{L1,L2,L3,L4,L5}  in the context of topological field theory.

\section{The lapse action, cusp caustics and Lefschetz thimbles}\label{Lefschetz}

Consider a general reparameterization invariant action,
\begin{align}\label{SpqN}
S[\boldsymbol Q(\tau),\boldsymbol P(\tau),N(\tau)] =
 \int d\tau \left( 
\boldsymbol P \cdot \dot{\boldsymbol Q} - N H(\boldsymbol P, \boldsymbol Q)
\right) \, .
\end{align} 
Reparameterizations are redefinitions of $\tau$ and $N$ under which $N(\tau)d\tau=N'(\tau')d\tau'$. The degree of freedom $N(\tau)$ is a Lagrange multiplier enforcing the Hamiltonian constraint,
\begin{align}
\frac{\delta}{\delta N}S = H = 0\, .
\end{align}
Among the examples of actions of this form are those which arise in mini-superspace reductions of gravitation, in which $N$ is referred to as the lapse and  $\hat H\Psi=0$ is the Wheeler-DeWitt equation. 
The Hamiltonian constraint is closely related to the existence of caustics for the following reason.  
Shadow zones, or domains of $\boldsymbol Q$ which cannot be  reached by real classical solutions from a given initial $\boldsymbol Q(0)$, may appear due to the constraint on initial velocities $\dot{\boldsymbol Q}(0)$ implied by $H=0$.  The boundaries of shadow zones are $A_2$ caustics, which are codimension $d=1$.  If these caustics terminate, they do so at caustics with $d\ge 2$.

The path integral representation \eqref{forml} of the wave function or a Green's function of $\hat H$,  
is strictly formal, requiring gauge fixing.  Details of the gauge fixing procedure can be found in  \cite{Halliwell,Teitelboim1,Teitelboim2}.  We chose to work in the gauge $\dot N=0$,  akin to a Schwinger proper time formalism in which $N$ is the proper time.   Integrating over $\boldsymbol P(\tau)$ and $\boldsymbol Q(\tau)$, with the boundaries $\boldsymbol Q(0)=\boldsymbol q$ and $\boldsymbol Q(1)=\boldsymbol q'$, gives
\begin{align}\label{psiN}
\Psi = \int_{\cal C}  dN \exp\left(\frac{i}{\hbar}{\mathbb S}
(\boldsymbol{q},\boldsymbol{q'},N)	\right)\, ,
\end{align} 
where we refer to $\mathbb S$ as the lapse action.
Whether $\Psi$ is a solution of $H(-i\hbar\nabla_{\boldsymbol q},\boldsymbol q)\Psi=0$ or a Green's function satisfying $ H(-i\hbar \nabla_{\boldsymbol q},\boldsymbol q)\Psi = \delta(\boldsymbol q-\boldsymbol q')$ depends on the choice of contour $\cal C$ in the complex $N$ plane.  
Since
\begin{align}\label{boundaryterms}
  H(-i\hbar\nabla_{\boldsymbol q},\boldsymbol q)\Psi = i\hbar\int_{\cal C} dN \partial_N\left(e^{\frac{i}{\hbar}\mathbb S} \right) =i\hbar\left. \exp\left( \frac{i}{\hbar}{\mathbb S}\right) \right|_{\partial {\cal C}}\, ,
\end{align}
one requires that $\exp(i{\mathbb S}/\hbar)$ vanishes as the boundaries of integration are approached,  with an exception for $\boldsymbol q'=\boldsymbol q$ in the case of a Green's function.   In general  $\exp(i{\mathbb S}/\hbar)$ has essential singularities at finite and infinite complex $N$ which can serve as boundaries of integration, provided they are approached within angular domains for which $\exp(i\mathbb S/\hbar)\rightarrow 0$.  

Assuming quadratic dependence on $\boldsymbol P$, path integration over $\boldsymbol P$ in \eqref{forml}  gives
\begin{align}
\Psi = \int dN e^{i\mathbb S}=\int dN \int{\cal D}\boldsymbol Q\, \exp \left[ \frac{i}{\hbar} \int d\tau\left (\frac{1}{N} \boldsymbol{\dot Q}^2  + NV(\boldsymbol Q) \right) \right]\,.
\end{align}
Integrating over $\boldsymbol Q$ in the small $N$ limit gives
\begin{align}\label{Nsmallsings}
\mathbb S = \frac{ (\boldsymbol q' - \boldsymbol q)^2}{4N} + \hbar \frac{d}{2}\ln(4\pi N) + \cdots\, .
\end{align}
Thus there is at least one finite pole of $\mathbb S$ at $N=0$.  
We shall set $\hbar =1$  in the subsequent discussion.  
Equation \eqref{boundaryterms} is that of a Greens function if the integration contour approaches $N=0$ on the negative imaginary axis, since
\begin{align}
\lim_{N\rightarrow -i0 } \exp(i{\mathbb S}) = \lim_{\epsilon\rightarrow 0} \frac{1}{ (4 \pi \epsilon)^{d/2}}\exp\left( -  \frac{ (\boldsymbol q'-\boldsymbol q)^2}{4\epsilon} \right) = \delta^d( \boldsymbol q' - \boldsymbol q)\, .
\end{align}
Obtaining a solution of $\hat H\Psi=0$ requires that the  integration contour does not contain a component ending at $N=0$, unless there is another beginning at $N=0$ such that a source and sink cancel.  In the latter case one can deform the contour away from $N=0$, although it will no longer be comprised of steepest descent paths, or Lefschetz thimbles. 

Singularities of ${\mathbb S}$ can also exist at finite non-zero $N$.  These were originally conjectured to  be related to the presence of non-smooth caustics in  \cite{G1,G2}.  The reason for the relation will be given in the subsequent section. A very simple example can be given for the Hamiltonian 
\begin{align}\label{trivham}
 H= -P_1^2 -P_2^2 + 1.
\end{align} 
Carrying out the path integration over $\boldsymbol P$ and $\boldsymbol Q$  with this Hamiltonian 
yields 
\begin{align}\label{ptsrc}
\exp\left({i\mathbb S}\right) =  \frac{1}{ 4 \pi N}\exp\left[i\left(\frac{ (q'_1 - q_1)^2 + (q_2'-q_2)^2}{4N} + N\right)\right]\, .
\end{align}
The Green's functions of $\hat H$ are those of a Helmholtz equation,
\begin{align}
\left( \partial_{q_1}^2 + \partial_{q_2}^2 + 1 \right) G(\boldsymbol q'|  \boldsymbol q) = \delta^2(\boldsymbol q'-\boldsymbol  q),
\end{align}
The lapse integral representation of the Green's function is  $G=\int dN \exp(i{\mathbb S})$, with complex contours connecting the essential singularities at $N=0$ and $N=+i\infty$.   Inequivalent contours connecting these singularities represent different boundary conditions for the Greens function.  
Thus far there are neither caustics nor finite singularities of the lapse action besides the pole at $N=0$.  However, despite the triviality of the Hamiltonian,   both $A_3$ caustics and finite $N$ poles of an effective lapse action exist for a suitable choice of source.

Consider a source which is a delta function in $q_1$ but smeared in $q_2$,
\begin{align}\label{linesrc}
\left(  \partial_{q_1}^2 + \partial_{q_2}^2 + 1 \right) \Psi = J(q_1,q_2) \equiv \delta(q_2 ) \sqrt{\frac{1}{4\pi\mu}} \exp \left( i\frac{1}{4\mu}q_1^2 \right)\, .
\end{align}
In this case
\begin{align}\label{splitpole}
\Psi &= \int JG = \int d{\bf q'}J({\bf q'})\int dN e^{i\mathbb S(\bf q,\bf q',N)}\nonumber\\ &=\int dN \frac{1}{4\pi}\sqrt{ \frac{1}{ N(N-\mu)}} \exp\left[i\left(\frac{q_2^2}{4N} + \frac{q_1^2}{4(N-\mu)} + N\right)\right]\, .
\end{align}
The integrand is similar to \eqref{ptsrc}, except that the $N=0$ pole of the exponent has been split in two. The exponent, an effective lapse action $\bar S$,  
has poles at $N=0$ and $N=\mu$.   An example will given later for which the lapse action has multiple finite poles due to the form of the Hamiltonian rather than the source. The current example is useful for its simplicity; it is a spherical cow model for an $A_3$ caustic in a diffeomorphism invariant theory. 

\begin{figure}[!h]
	\center{
		\includegraphics[width= 350pt]{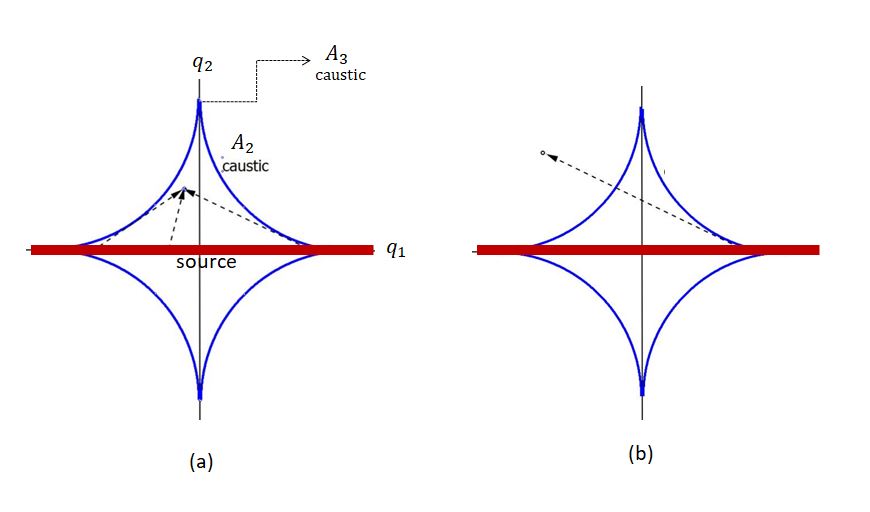}
		\caption{The caustics associated with the lapse action \ref{cusplapse} form an astroid with cusp ($A_3$) caustics at the corners connected by smooth ($A_2$) caustics.   For any point within the astroid, there are three real critical points of the lapse action, with the corresponding rays solving \eqref{rays} and the source condition \eqref{raysbc}  shown in (a).   For a point outside the astroid, there is a single real critical point, with the corresponding ray shown in (b). } 
		\label{astroid}
	}\end{figure}

The critical points of the effective lapse action are $N=N_c$ satisfying  $\frac{\partial \bar S}{\partial N} =0$,  where\footnote{For simplicity we have not lifted the factor of $[N(N-\mu)]^{-1/2}$ into logarithmic terms in the definition of the modified lapse action $\bar S$. 
The logarithmic terms vanish in  the classical $\hbar\rightarrow 0$ limit, as in \eqref{Nsmallsings}, and have no bearing on the geometry of the caustics.  }
\begin{align}\label{cusplapse}
\bar S \equiv \frac{q_2^2}{4N} + \frac{q_1^2}{4(N-\mu)} + N\,  .
\end{align}
 Critical points coalesce in pairs along the smooth curves of the astroid,
\begin{align}\label{astrd} 
q_1^{2/3}+q_2^{2/3} = (2\mu)^{2/3}\, ,
\end{align}
shown in figure \eqref{astroid}, except at the cusps, such as $q_1=0,q_2=\pm 2\mu$, at which a triplet of critical points merge. The smooth curves of the astroid are $A_2$ caustics, while the cusp points are $A_3$ caustics.  Points $(q_1,q_2)$ in the neighborhood of a cusp corresponds to unfoldings of the $A_3$ caustic.

In the vicinity of a cusp, there is a locally defined
map $(N,q_1,q_2) \leftrightarrow (\lambda,\zeta_1,\zeta_2)$
such that $dN \exp(i\bar S) \sim d\lambda \exp(i{\cal P})$ where
${\cal P}$ is the potential function for an $A_3$ catastrophe,
\begin{align}\label{A3poly}
{\cal P} = \lambda^4 + \zeta_2\lambda^2 + \zeta_1\lambda\, .
\end{align}
The caustic curve associated with \eqref{A3poly} along which $\frac{dP}{d\lambda}=0$ and $\frac{d^2P}{d\lambda} = 0$, is given by
\begin{align}
8\zeta_2^3 + 27\zeta_1^2 = 0\, ,
\end{align}
which describes any corner of the astroid \eqref{astrd}.  The mapping of the region in the neighborhood of $q_1=0$ and $q_2 = 2\mu$ is given by 
\begin{align}\label{astroidmap}
\zeta_2 &=  \gamma(2\mu)^{-1/3}(q_2 - 2\mu)\nonumber \\
\zeta_1 &=  \gamma^{3/2} q_1 \nonumber \\
\gamma &= 2^{1/3}\mu^{-1/6}
\end{align}
The map relating $N$ and $\lambda$ has been partially described in \cite{G2}, and is singular along the curves $q_1 = 0$ and $\zeta_2=0$ which intersect at the cusp.
There are multiple cusp caustics associated with the lapse action $\bar S(N)$ but only one with a potential function ${\cal P}(\lambda)$, so no map can be given globally.  Near a particular cusp caustic, the wave function is described approximately by a Pearcey function,
\begin{align}
\Psi(q_1,q_2) \approx \int d\lambda e^{i(\lambda^4 + \zeta_2\lambda^2 + \zeta_1\lambda)}
\end{align}


\begin{figure}[!h]\label{splitpolethimbles}
	\center{
		\includegraphics[width= 350pt]{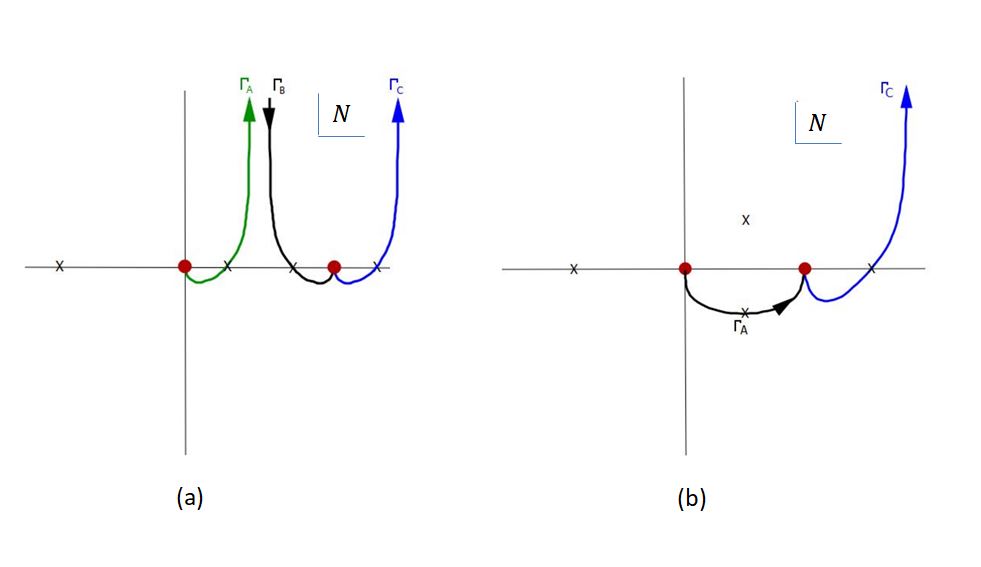}
		\caption{An example of Lefschetz thimbles in the complex lapse plane, associated with a solution of \eqref{linesrc} with radiation boundary condition. The finite $N$ poles of the lapse action, shown as circles, are endpoints of thimbles, each of which passes through a critical point shown as an X.  In (a) $\bf q$ is inside  the  caustic astroid and the thimbles include three real critical points.  Moving  $\bf q$ across the caustic gives the thimbles  shown in (b), including one real and one complex critical point. In both cases there are two thimbles ending at the pole $N=\mu$, such that the locus of vanishing residue $q_1=0$ is not a source, but a curve to which the cusp points are bound. }\label{splitpolethimbles}
		
	}\end{figure}

Caustics are often defined in terms of the behavior of groups of rays, i.e. the curves transverse to wave-fronts. These are equivalent to trajectories  satisfying the classical equations of motion,
\begin{align}\label{rays}
\dot {\boldsymbol q} &= \{ {\boldsymbol q},H \} \nonumber \\
 \dot {\boldsymbol p} &= \{ {\boldsymbol p},H \} \nonumber \\ 
H&(\boldsymbol{q}, \boldsymbol{p})=0\, .
\end{align}
 In the present example, with the Hamiltonian  \eqref{trivham} and the line source \eqref{linesrc}, the rays are straight lines emanating from the source at $q_2=0$ in the direction
\begin{align}\label{raysbc}
 [\hat n_1,\hat n_2] =    \left[-\frac{q_1}{2\mu}, \sqrt{1-\left( \frac{q_1}{2\mu}\right)^2}\, \right]\, .
\end{align}
These are equivalent to flows generated by $\nabla_{\bf q} \bar S(q_1	,q_2,N_c)$. 
There are three rays reaching any point within the astroid, but only one (real) ray reaching any outside point. The density of a bundle of rays diverges as one approaches the astroid. 

There is a steepest descent path, or Lefschetz thimble, associated with each critical point of $\bar S(N)$.
Lefschetz thimbles associated with \eqref{cusplapse} and radiation boundary conditions are shown in figure \ref{splitpolethimbles}, for points both inside and outside the astroid.  Inside the astroid there are three thimbles passing through the three real critical points. Outside the astroid there are two thimbles, passing through a single real critical point and a single complex critical point.
Whether inside or outside the astroid, one of the contributing Lefschetz thimbles terminates at the $N=0$ pole of $\bar S$,  corresponding to the line source at the locus of vanishing residue $q_2=0$.  On the other hand one thimble ends and another begins at the $N= \mu$ pole of $\bar S$, such that there is no source term at the locus of vanishing residue $q_1=0$.   Nevertheless, the line $q_1=0$, dubbed a ghost source in \cite{G1}, is physical in the sense that the cusps are bound to it.  The relation between caustics with codimension $d\ge 2$, ghost sources and the finite $N$ singularities of the lapse action will elaborated upon  in section \ref{polesandcaustics}.

\section{Relating  finite $N$ singularities of the lapse action to $A_{n\ge 3}$ caustics}\label{polesandcaustics}

The simultaneous presence of finite $N$ poles in the lapse action and cusp caustics
is a general phenomenon for reasons we now discuss.
Since the lapse action is given by 
$\exp(i\mathbb S)=\int {\cal D}{\bf P}{\cal D}{\bf Q}exp(i[{\cal L}({\bf Q,\bf P},N)]$, consider the equations of motion associated with the Lagrangian $\cal L$,
where $\bf Q$ and $\bf P$ are variational parameters but not $N$. 
For certain values of the lapse, $N=N_p$, the Euler-Lagrange equations may only have solutions for a collapsed set of boundary conditions.   
An example will be given in the next section 
involving a  
a minisuperspace model for which classical solutions with the same ${\bf Q}(0)=q'$, distinguished by different $\bf P(0)$,
all have the same endpoint ${\bf Q}(1)=q$.  The absence of classical solutions for any other endpoint is reflected in a divergence of ${\mathbb S}(N)$ as $N\rightarrow N_p$ for  disallowed boundary conditions.  We will find singularities of ${\mathbb S}$ of the form 
\begin{align}
{\mathbb S} (N) \sim  \frac{   f(\bf q,\bf q')}{ N-N_p }\, ,
\end{align}
where the residue of the pole, $f(\bf q,\bf q')$, vanishes for  the classically allowed boundary conditions at $N=N_p$. As discussed in section \ref{Lefschetz}, the 
locus of vanishing residue is  either a source or a ghost source, depending on the combination of  Lefschetz thimbles with  endpoints at $N=N_p$.  

Treating $N$ as a parameter, the classical solutions are an extended set which need not satisfy the Hamiltonian constraint $H(\bf P,\bf Q)=0$. For the constraint to hold, $N$ must be a critical point $N_c$ of the lapse action.
The collapse of the allowed boundary conditions for the extended solution set as $N\rightarrow N_p$ suggests that critical points in the neighborhood of $N_p$ may have non-generic physical properties.  

\begin{figure}[!h]
	\center{
		\includegraphics[width= 300pt]{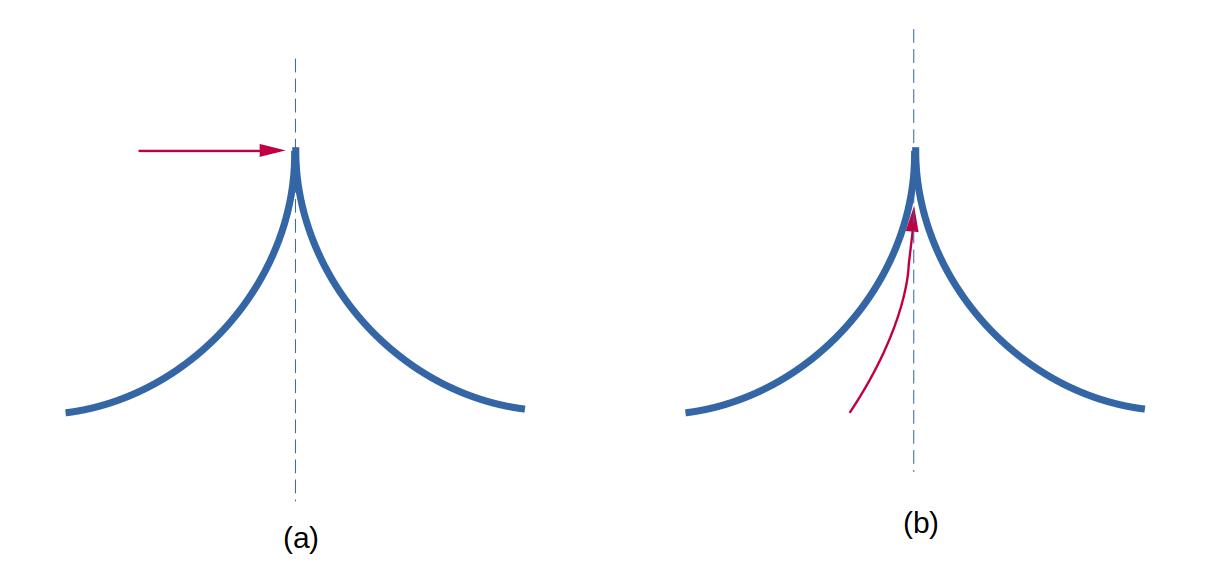}
		\caption{(a) Approaching a cusp caustic via the coalescence of two complex and one real critical point. (b) Approaching a cusp caustic via the coalescence of three real critical points. }  \label{Approach}}
\end{figure}

To investigate the properties of critical points near a pole, consider a Laurent expansion of the lapse action about a simple pole,
\begin{align} \label{Laurent}
{\mathbb S}\approx &\sum_{n=-1}^\infty g_n (N-N_p)^n \\ 
&g_{-1} = f(\bf q, \bf q') \sim \xi^2 \, , \label{resdu}
\end{align}
where $\xi$ parameterizes the distance to the locus of vanishing residue, or ghost source.
Critical points in the neighborhood of the pole satisfy
\begin{align}\label{critpole}
\partial_N {\mathbb S} =-\frac{\xi^2}{ (N_c-N_p)^2} + g_1 + g_2 (N_c-N_p) + \cdots =0
\end{align}
Solutions of \eqref{critpole} in the limit $N_c-N_p \rightarrow 0$ 
satisfy a scaling relation 
\begin{align}\label{scal}
\xi^2 \sim(N_c-N_p)^{\nu}\, ,
\end{align}
where the index $\nu$  depends on the lowest $n=n_{bottom}\ge 1$ for which $g_n$ in \eqref{critpole} does not vanish:  
\begin{align}
\nu = n_{bottom}+1\, .
\end{align}
There are coalescing critical points as $\xi\rightarrow 0$, the number of which is determined by $\nu$.
If  $n_{bottom}=2$,  $N_c-N_p = \xi^{2/3}$ and there are three coalescing critical points indicating an $A_3$ caustic.  The lapse action at the critical points behaves as ${\mathbb S}_c \approx {\mathbb S}_0 + c \xi^{4/3}$,  which maps to the behavior of the potential function \eqref{A3poly}  with  $\zeta_2 \sim \xi$ and $\zeta_1=0$.  The corresponding approach to the cusp point is shown in figure \ref{Approach}a.
Thus one finds $A_3$ caustics at the intersection of the surface  $g_1(\bf q,\bf q')=0$, corresponding to $n_{bottom}=2$,  with the ghost source\footnote{These are the same surfaces along which the map $\lambda(N)$ is singular.}  $g_{-1}(\bf q,\bf q')=0$.   This localization will be shown explicitly in the example of section \ref{examples},  in which there are multiple $A_3$ caustics bound to ghost sources.  
The case $n_{bottom}=1$ is subtle; it does not indicate an $A_2$ caustic even though two critical points collide, for reasons discussed below.

\subsection{Spurious collisions and singularities of the map $N\leftrightarrow\lambda$}

There is a complicating subtlety in relating collisions of critical points in the complex lapse plane to caustics.  The problem is the existence of collisions of $N_c$ at ghost sources which are spurious, in the sense that there is no map from the lapse action $\mathbb S(N)$ to a  potential function $P(\lambda)$ which preserves the collision. 
The collision is an artifact of singularities in the map $\lambda \leftrightarrow N$ at a ghost source.   

An example of a collision of critical points in the complex $N$ plane which is unrelated to caustics occurs at a generic point along a ghost source, at which $g_{-1}=0, g_1 \ne 0$ and $n_{bottom} =1$. Here there are two coalescing critical points $N_c-N_p \sim \pm \xi$ with ${\mathbb S} \approx {\mathbb S}_0 \pm c\xi$.  This behavior reflect a degeneracy but not an $A_2$ caustic.  
For an $A_2$ caustic the potential function ${\cal P}=\lambda^3 + \zeta_1\lambda$ becomes ${\cal P} \sim \zeta_1^{3/2}$ at the critical points. The two critical points transition from real on one side of the caustic to complex on the other.
In the language of ray theory,  there is a pair of real rays in the `illuminated zone' which become complex in the `shadow zone'. However the pair of real critical points of ${\mathbb S}(N)$  which collide at a generic point along the ghost source simply pass through each other while remaining real and the lapse action is analytic in $\xi$. There is no such collision after mapping $N\rightarrow\lambda$, reflecting the singularity of the map $\lambda(N)$ at a ghost source.  Indeed, the actual $A_2$ caustics do not lie on the ghost source, at which one can find only their boundary: a higher order caustic. 
    
\begin{figure}[!h]
	\center{
		\includegraphics[width= 200pt]{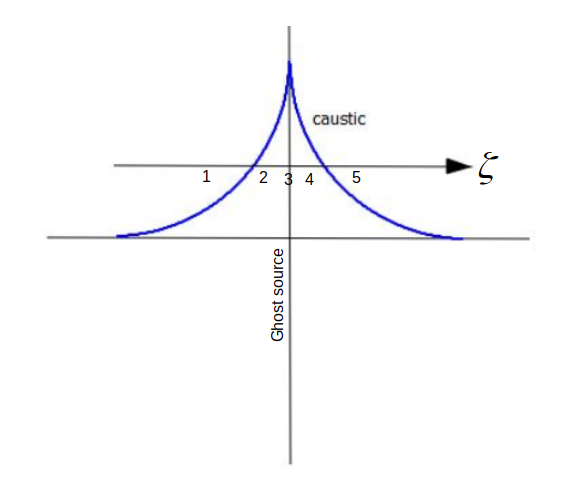}
		\caption{Traversing the neighborhood of a cusp caustic, bisected by a ghost source along which the residue of a pole of the lapse action vanishes.} \label{cuspcross}}  
\end{figure}
\begin{figure}[!h]
	\center{
		\includegraphics[width= 350pt]{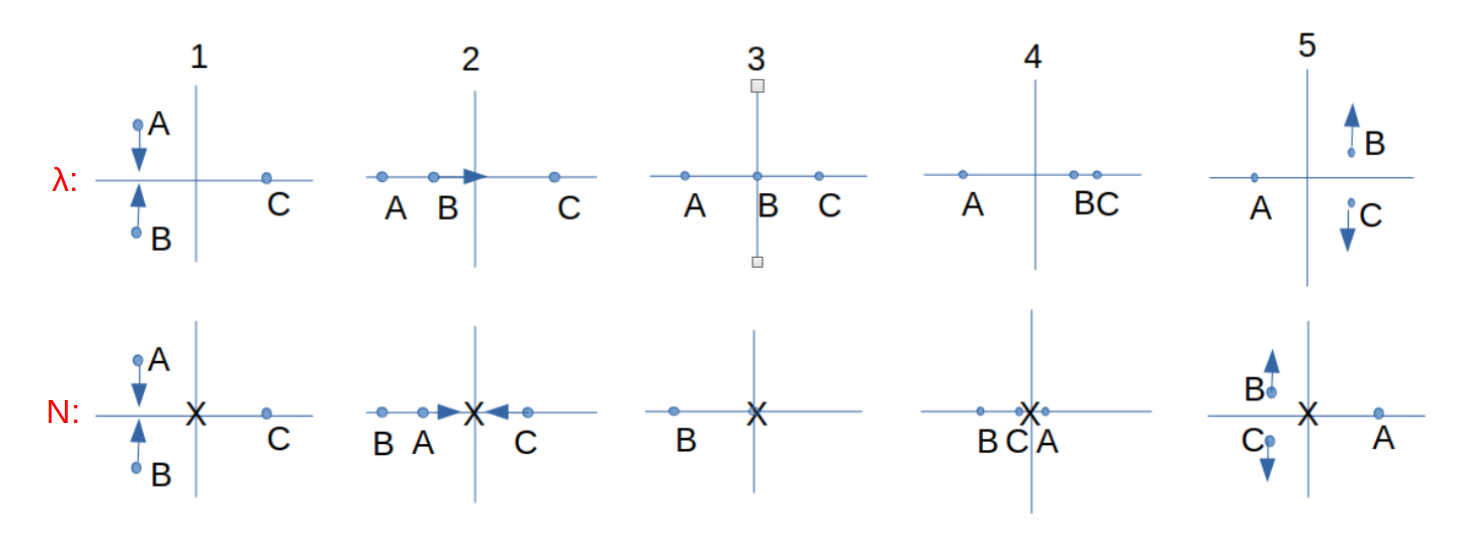}
		\caption{The first row shows the evolution of critical points of the potential function \eqref{A3poly}  in the complex $\lambda$ plane during a crossing of the cusp caustic shown in figure \ref{cuspcross}.  The second row shows the same evolution for the critical points of the lapse action \eqref{cusplapse} in the complex $N$ plane, in which the $N=\mu$ pole is indicated by $X$.  The crossing of the ghost source occurs in the third column. In the complex $N$ plane, the two critical points $A$ and $C$ cross through each other at the ghost source,  however nothing dramatic, such as an $A_2$ caustic exists at this point: there is no such coincidence of critical points in the complex $\lambda$ plane.  The map between $N$ and $\lambda$ is singular along the ghost source. } \label{fk}}
\end{figure}

It is enlightening to compare the behavior of the critical points of the potential function to the critical points of the lapse action for a crossing of the astroid near a cusp, as shown in figure \ref{cuspcross},  in the context of the simple example described in section \ref{Lefschetz}. 
The correspondence between the critical points is obtained using the map \eqref{astroidmap}, yielding the
behavior shown in figure \ref{fk}. 
Instead of merging to form a higher order critical point, the  pair of critical points of ${\mathbb S}(N)$ which converge on $N_p$ as the the ghost source is approached, cease to exist upon reaching $N_p$.  At this point the square root terms proceeding the exponential in \eqref{splitpole}, which we have so far been ignoring, become very important;  the relative homology is preserved by the presence of the branch cuts and contours which begin and end at infinity on different Riemann sheets (see \cite{G1}).


Explicit examples with $n_{bottom} \ge 3$, which could map to $A_{n\ge 4}$ caustics, have yet to be constructed.  Here too, one would have to be careful to determine whether collisions of critical points at the ghost source are caustics or artifacts of the singular map between $N$ and $\lambda$. 


\subsection{Caustics and the Riemann Hurwitz formula}

It is not obvious that $A_3$ caustics can only appear along ghost sources.  
Using the Riemann-Hurwitz formula, one can construct  an argument for the case in which a few assumptions are satisfied.  The formula gives an association between poles and critical points such that
three critical points can not collide without a pole getting in the way;  therefore, one must simultaneously approach a ghost source, $g_{-1}\rightarrow 0$.

Consider ${\mathbb S}(N)$ as a map of a Riemann surface of Euler characteristic $\chi$ onto another of Euler characteristic $\chi'$. For a degree $\beta$ covering map, the Riemann-Hurwitz formula is,
\begin{align}\label{RH}
\chi=\beta\chi' - \sum_i(m_i-1),
\end{align}
where the integer $m_i$ is the ramification index at the i'th branch point of the inverse map $N({\mathbb S})$.   
Assuming large $N$ asymptotics of the form $S\sim N^{\ell_{top}}$, the ramification index at infinity is $m_\infty = \ell_{top}$.
At the $i$'th critical point ${\mathbb S}\sim (N-N_{c,i})^{m_i}$, with $m_i\ge 2$. Caustics correspond to $m_i \ge 3$. Away from caustics $\sum_{i,\rm finite} (m_i-1)$ is simply the number of critical points $n_{c}$,
If ${\mathbb S}(N)$ is  meromorphic, then the degree $\beta$ of the map is the sum of the order of the poles, including that at infinity.  For $n_P$ finite simple poles,
\begin{align}
\beta = m_\infty + n_P = \ell_{top}+n_P\, .
\end{align}
Assuming a map between Riemann spheres,  $\chi=\chi'=2$, the Riemann--Hurwitz formula \eqref{RH} becomes 
\begin{align}\label{gnrl}
n_c = 2n_P + \ell_{top} -1 \, .
\end{align}

With the redefinition $\tau'=N\tau$,  
\begin{align}\label{asyN}
\int_0^1 d\tau {\cal L} = \int_0^N d\tau' \left( {\bf P \cdot \dot Q - H(\bf P,\bf Q)}  \right)\sim N
\end{align} 
at large $N$,  assuming that the mean value of ${\bf P} \cdot \dot {\bf  Q}-H(\bf P,\bf Q)$  along an arbitrarily long  (large $N$) classical trajectory is bounded.  
For the asymptotics \eqref{asyN},  $\ell_{top} = 1$ and
\begin{align}\label{cp}
n_c=2n_P\,.
\end{align}
While \eqref{asyN} holds in the explicit examples considered here,  we caution  that the classical trajectories are not necessarily bounded; there are examples of quantum cosmologies such as those considered in \cite{TurokLorentzian}, for which \eqref{gnrl} still applies but $\ell_{top}>1$.  However none of these models have either codimension $d\ge 2$ caustics or finite poles beside $N=0$.
If based on \eqref{cp} one makes the further assumption that real critical points arrange themselves in pairs separated by real poles, except for a single critical point to the left of the leftmost pole and conversely on the right,
then the collision of three real critical points
can only occur by converging on a pole.  
All $A_3$ caustics are then bound to ghost sources. 

The case for this ordering of poles and critical points can be made presuming the existence of a set of Lefschetz thimbles which is equivalent to the real axis in the Lorentzian path integral \cite{TurokLorentzian}.
Each Lefschetz thimble connects a pair of poles while passing through a critical point; the set equivalent to the real axis can be written as  
\begin{align}\label{Lorpath}
{\cal C} = \{\infty \rightarrow N_{1}\}, \{N_{1}\rightarrow N_{2}\}, \cdots \{N_{n_P -1}\rightarrow N_{n_P}\}, \{N_{n_P}\rightarrow\infty\}\, ,
\end{align} 
where $\infty$ refers to the simple pole at infinity.  Note that the connecting path between any pair of adjacent finite poles in this notation, $\{N_i \rightarrow N_{i+1}\}$ can be a single Lefschetz thimble or two Lefschetz thimbles, each with a boundary at $\infty$; 
\begin{align}
\{N_{i}\rightarrow N_{i+1}\} = \{N_{i}\rightarrow\infty\},\{\infty\rightarrow N_{i+1}\}\, .
\end{align}
These two possibilities are separated by an $A_2$ caustic;  there are two real critical points on one side of an $A_2$ caustic, whereas there is only a single complex critical point which contributes on the other side\footnote{There is also a complex conjugate critical point which does not lie on the set of Lefschetz thimbles  equivalent to the real axis.}.  
Combining the structure \eqref{Lorpath} with the fact that Lefschetz thimbles can not intersect forces the real critical points to be arranged in the way we have described, i.e. pairs between poles. This arrangement of poles and critical points is observed in the simple model of section \ref{Lefschetz} as well as in the more complicated model to be described in section \ref{examples}.  Although the latter exhibits an infinite number of poles and critical points, so that \eqref{cp} is ill defined, real critical points still occur in pairs between poles.

An exception to the arrangement we have described here could occur if the pole at infinity is not simple, or $\ell_{top}>1$.  In this case \eqref{cp} would be modified and there would be more than one angular domain in which  the lapse integration contour could approach infinity.  For example if $\ell_{top} =2$, there are two such domains $\infty_1$ and $\infty_2$, and one could have 
\begin{align}
\{N_{i}\rightarrow N_{i+1}\} = \{N_i\rightarrow\infty_1\},\{\infty_1\rightarrow\infty_2\},\{\infty_2\rightarrow N_{i+1}\}\, ,
\end{align} 
allowing three real critical points between a pair of poles.
The asymptotics \eqref{asyN} renders this impossible for bounded classical evolution.  For an unbounded case with $\ell_{top}>1$, it is not clear that all $A_{3}$ caustics are bound to ghost sources.  However higher order $A_{n \ge \ell_{top}+2}$ caustics, assuming their existence, would necessarily be bound to ghost sources by an argument analogous to the one we have just given. 
  



\section{Cusp caustics in the wave function of\\ the universe} \label{examples}

A minisuperspace model of a (2+1) dimensional universe considered by Halliwell and Myers in \cite{HalliwellMyers}
can be modified in a simple way to illustrate the relation between finite $N$ poles of the lapse action and $A_3$ caustics. 
The lapse action in this case has an infinite number of poles.   However, in the original model, the wave function depends solely on a scale factor and therefore has no caustics of codimension greater than one. We will construct a variant with  an additional scalar field, akin to a clock variable, in which there are $A_3$ caustics in the scale factor -- clock plane.  The cusps are located at points along ghost sources predicted by the arguments of section \ref{polesandcaustics}. 
%



The action considered in \cite{HalliwellMyers} is that of (2+1)  dimensional Euclidean Einstein gravity with a positive cosmological constant $\Lambda$, a Gibbons-Hawking boundary term and a rank 2 antisymmetric tensor matter field $A$,
\begin{align}\label{actn}
I=-\frac{1}{16\pi G}\int d^3  x\, g^{1/2}\, (R-2\Lambda) - \frac{1}{8\pi G} \int d^2 x\, h^{1/2} K + \int d^3x\, g^{1/2} A^2\, .
\end{align}
The degrees of freedom are truncated in minisuperspace, restricting the metric to the form
\begin{align}
ds^2 = \frac{G^2}{16\pi^2}(N(\tau)^2d\tau^2 +{\tiny {\tiny }} a(\tau)^2d\Omega_2^2)\, ,
\end{align}
where $d\Omega_2^2$ is the metric of the unit 2-sphere.  The tensor field is taken to be $A=M\epsilon$ where $\epsilon$ is the 2-dimensional volume form on a constant $\tau$ surface, normalized so $\int \epsilon=1$. The quantity $M$ is the tensor charge, chosen to vanish in the solutions considered here.  
The Euclidean minisuperspace action is
\begin{align}\label{Naction}
I= \frac{1}{2} \int_0^1 &d\tau \left[i\pi_a \dot a -NH(\pi_a,a)\right]\nonumber \\
&H \equiv -\pi_a^2 +1 - \lambda a^2\, ,
\end{align}
with $\lambda\equiv \Lambda\frac{G^2}{16\pi^2}$. Equivalently, integrating out $\pi_a$,
\begin{align}\label{narc}
I = \frac{1}{2} \int_0^1 d\tau\,\left[-\frac{  {\dot a}^2 }{N} - N(1-\lambda a^2)   \right]\, .
\end{align}  
Solutions of the equations of motion oscillate about $a=0$, with $|a|$ corresponding to the two-sphere radius. The corresponding geometry consists of a pair of two spheres of radius $a(0)$ and $a(1)$ at the boundary of three-balls, which are connected via multiple  intermediate contiguous three spheres.  The connection points are wormholes at which $a=0$.

Choosing the gauge $\dot N=0$, 
the  propagator is given by the path integral,
\begin{align}\label{scalp}
 G(a'|a'') = \int_{\cal C} dN \int {\cal D}a\exp\left(  
-\frac{1}{2} \int_0^1 d\tau\,\left[-\frac{  {\dot a}^2 }{N} - N(1-\lambda a^2)   \right]
 \right)
\end{align}
with $a(0)=a', a(1) = a''$.
If the complex contour ${\cal C}$  is bounded at $N=0$ and infinity,  \eqref{scalp}  
is equivalent to a Greens function of the Hamiltonian operator,
\begin{align}
\hat H=\partial_{a''}^2 + (1-\lambda {a''}^2), 
\end{align}
satisfying $\hat H G = \delta(a''-a')$.
On the other hand if the contour has no endpoints or can be deformed such that it only has endpoints at infinity, then 
$\hat H G=0$ and the wave function $\Psi(a'') \equiv G(a'|a'')$ is a solution of the Wheeler-DeWitt equation.
Setting $a'=0$ yields a wave function of the universe  satisfying  Hartle-Hawking boundary conditions \cite{HH}.  
There are a multitude of Green's functions of $\hat H$ and solutions of the Wheeler-DeWitt equation $\hat H\Psi=0$, associated with different sums over complex $N$ contours in either a Euclidean or Lorentzian path integral.  The case for the Lorentzian formulation was made in \cite{TurokLorentzian}, based on the unique deformation of a conditionally convergent integration of $\exp(i{\mathbb S})$ over the real axis to a sum of absolutely convergent integrals over Lefshetz thimbles. For present purposes, we are interested in the singularities in the complex $N$ plane and their connection to non-smooth caustics, rather than 
the physical justification for choosing particular contours in a Euclidean or Lorentzian path integral.    

Regarding $N$ as a parameter without imposing the Hamiltonian constraint, the equation of motion for the action \eqref{narc} 
is that of harmonic oscillator,
\begin{align}\label{harmon}
\ddot a+N^2\lambda a =0\, .
\end{align}
At certain values of $N$, solutions do not exist for all boundary values $a'$ and $a''$.
For $N=\frac{\pi n}{\sqrt{\lambda}}$ with integer $n$, 
\begin{align}\label{allowed}
a'' - (-1)^n a'=0\, ,
\end{align}
regardless of the intial velocity $\dot a(0)$.
The collapse of allowed boundary conditions is reflected in the singularities of the lapse action ${\mathbb S}(N)$.  Carrying out the path integration over $a(\tau)$ yields
\begin{align}\label{H1}
e^{-  {\mathbb S}(N)  } &\equiv \int {\cal D}a \exp(-I[a(\tau),N]) \nonumber \\
&=  \frac{1}{(sin(\lambda^{1/2}N))^{1/2}} \exp\left[   
	-\frac{1}{2} 
		\left( 
			N + \lambda^{1/2}\frac{(a'^2 + a''^2)cos(\lambda^{1/2}N) - 2 a' a'' }{sin(\lambda^{1/2}N)}
		\right)
			\right]\, .
\end{align}
Thus $\mathbb S$ has poles and coincident logarithmic branch points at $N=\frac{\pi n}{\sqrt{\lambda}}$.
The residues of the poles vanish for the allowed boundary conditions \eqref{allowed}.  The locus of vanishing residue is interpreted as either a source or a ghost source depending on the choice of contours included in the lapse integral. 

Greens functions and solutions of the Wheeler--DeWitt equation can be represented as sums of integrals over 
Lefschetz thimbles connecting essential singularities of $\exp(-{\mathbb S}(N))$, each of which passes through critical points corresponding to the wormhole connected three geometries described above. 
The essential singularities of \eqref{H1} are the   finite $N$ simple poles of ${\mathbb S}(N)$ as well as a simple pole at infinity, at which ${\mathbb S}\sim N$. 

\begin{figure}[!h]
	\center{ 
		\includegraphics[width= 420pt]{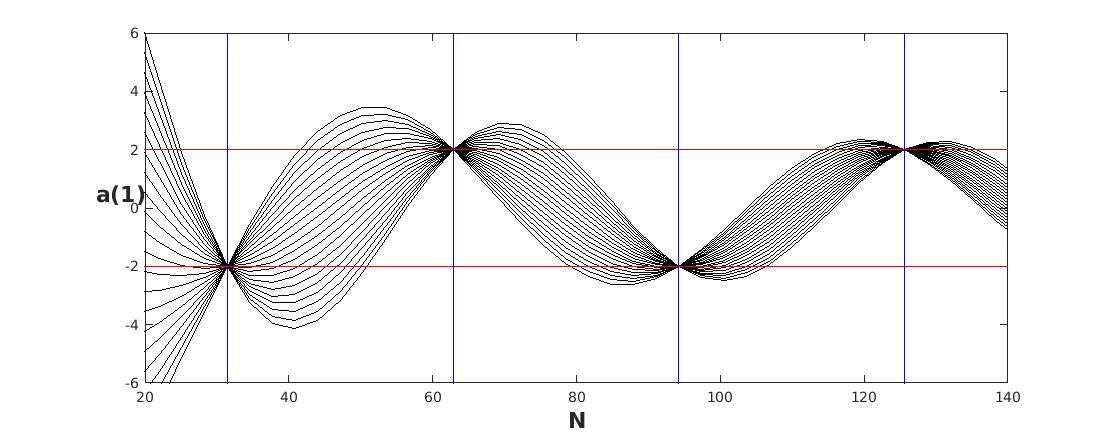}
		\caption{A fan of trajectories showing the endpoint $a''\equiv a(\tau=1)$ as a function of the lapse $N$ for $\lambda=0.01$, fixed initial $a(\tau=0)=2$ and a range of initial velocities $\dot a(\tau=0)$.  The possible endpoints collapse at $N=\frac{\pi m}{\sqrt{\lambda}}$, which are the poles of $\mathbb S$, to a value given by setting the residue to zero.  The poles and vanishing residues correspond to the vertical and horizontal lines respectively.   
		.}
		\label{fan1}
	}\end{figure}

As noted above, the wave function is only defined over one degree of freedom 
such that there are no caustics with codimension greater than one.  However a precursor of $A_3$ caustics  can be seen as follows.  Consider a fan of trajectories $a''(N)$, or boundaries $a(\tau =1)$ as a function of the lapse,  defined by the equations of motion \eqref{harmon} with a fixed $a'=a(\tau=0)$ and a range of $\dot a(\tau = 0)$. 
The fan shows focal regions at values of $N$ where the solution set collapses, as illustrated in figure \ref{fan1}. These points lie at the intersection of the poles of ${\mathbb S}(N)$ and the values of $a''$ at which the residues vanish.  When a scalar field $T$ acting as a physical clock is added to the model, we will see that there is a one to one correspondence between $A_3$ caustics in the $a,T$ plane and the focal points seen in figure \ref{fan1}.

\begin{figure}[!h]
	\center{
		\includegraphics[width= 400pt]{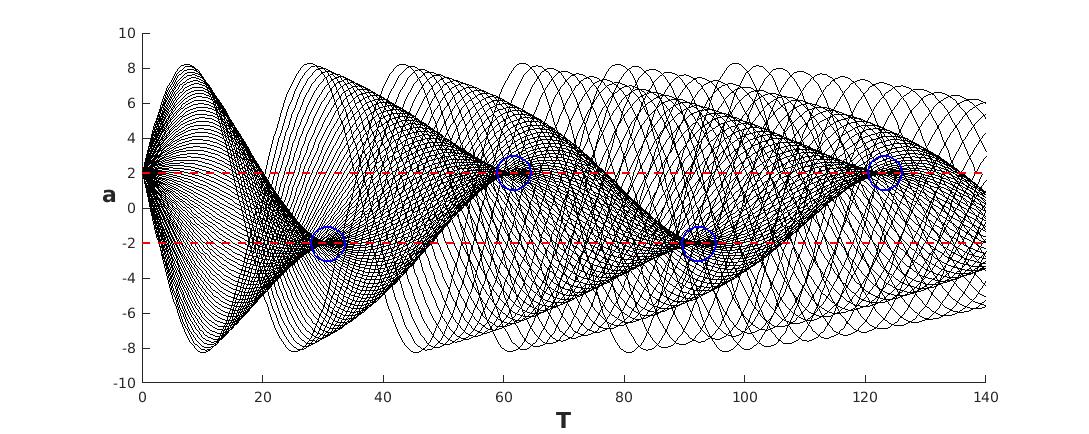}
		\caption{A collection of classical trajectories $a(T)$ for initial condition $a(0)=2$ and $\lambda=0.01$, showing a sequence of cusp caustics (encircled).  The cusp caustics are bound to ghost sources,  shown as horizontal lines in the figure, at which the residues of poles of ${\mathbb S}(N)$ vanish, $g_{-1}=0$.  Each cusp is associated with a particular pole $N_m$, with location along the ghost source determined by setting $g_1=0$, where $g_1$ is the order $N-N_m$ term in the Laurent expansion of ${\mathbb S}$ about the pole.}
		\label{cuspchannel1}
	}\end{figure}

\begin{figure}[!h]
	\center{
		\includegraphics[width= 250pt]{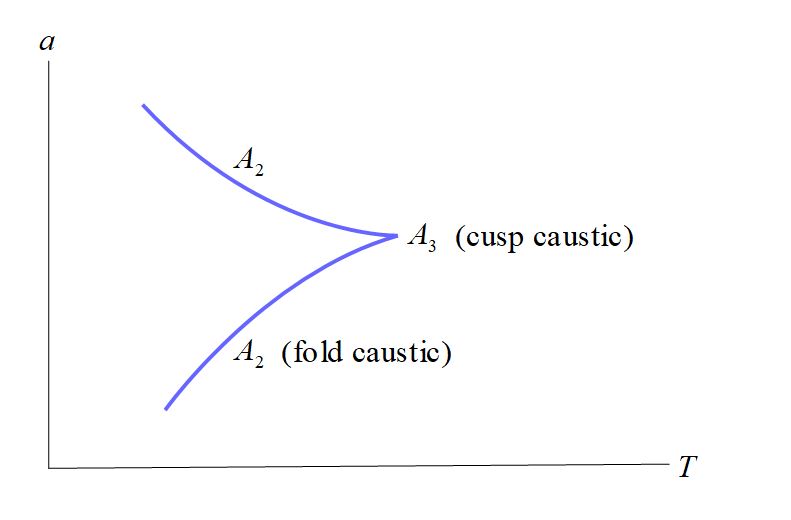}
		\caption{An illustration of a cusp caustic, which is the codimension two intersection of a pair of  fold caustics. .}
		\label{cuspillust}
	}\end{figure}

We now add a clock variable  to the model \eqref{actn}, in the form of a massless scalar field $T$ with a `wrong sign'  kinetic term.
 \begin{align}\label{actnplus}
I=-&\frac{1}{16\pi G}\int d^3 \, x g^{1/2}\, (R-2\Lambda) - \frac{1}{8\pi G} \int d^2 x\, h^{1/2} K + \int d^3x\, g^{1/2} A^2 \nonumber\\
&- \frac{1}{2}\int d^3 x g^{1/2} (\partial T)^2\, .
\end{align}
In the minisuperspace truncation, the spatial dependence of $T$ is dropped, giving the action
\begin{align}\label{Nsup}
I=\int_0^1 d\tau \left[ \frac{1}{2N}(-{\dot T}^2 - {\dot a}^2) - \frac{N}{2}(1-\lambda a^2)  \right]\, .
\end{align}
The equation of motion for $T$ is $\ddot T=0$, with solution $T=\gamma \tau$ known as a ghost\footnote{The nomenclature is un-related to the ``ghost sources.''} condensate \cite{ArkaniHamed}.  The only coupling between $T$ and $a$ is due to the Hamiltonian constraint. An integral form of the constraint, $\int d\tau H=0$, follows from $\frac{dI}{dN}=0$, but the integral can be dropped by conservation of $H$. Figure \ref{cuspchannel1} shows a plot of a collection of rays, or the curves $a(T)$ obtained by solving the equations of motion\footnote{The model described here is coincidentally very similar one appearing in an entirely different physical context; an ocean acoustic sound channel \cite{Holford}.   Mapping $T$ to a horizontal coordinate $X$, $a$ to a depth coordinate $Z$, and replacing $\exp(-I[a(\tau),T(\tau),N])$ with $\exp(iI(Z(\tau),X(\tau),N))$ yields a path integral solution of the Helmholtz equation in a sound channel with index of refraction $n(\vec x) = 1 - Z^2$.  } including the Hamiltonian constraint, for a particular initial $a(0)=a', T(0)=0$. 
The set of rays can be parameterized by $\sigma \equiv \dot a(0)$.  The density of rays, or the Jacobian 
$\frac{ \partial(\sigma,\tau)}{\partial(a,T)}$,
diverges at caustics. The caustics in figure \ref{cuspchannel1} have the basic structure depicted in figure \ref{cuspillust}, in which 
pairs of codimension one $A_2$ caustics end at codimension two $A_3$  caustics. The wave function is large near the $A_2$ caustics, taking its largest values in the neighborhood of their $A_3$ terminus.  	
However because the path integral is Euclidean,  the wave function in the neighborhood of the $A_3$ caustics is not the usual diffraction catastrophe, but an analytic continuation of the Pearcey function to complex arguments.    

Let us now consider the singularity structure of the lapse action derived from the model \eqref{Nsup} which includes the clock variable.
The propagator is
\begin{align}\label{sbarchan}
G(a',T'|a'',T'')&=\int dN \int {\cal D}a {\cal D}T 
e^{-I[a,T,N]} \nonumber\\
&=\int dN \frac{1}{4\pi  N}[{\rm sinc}(\lambda^{1/2}N)]^{-1/2} \exp(-S(N))\, , \nonumber \\
S(N) = &-\frac{1}{2}\left(\frac{ (T''-T')^2}{2N} + N + \lambda^{1/2}\frac{ (a''^2 + a'^2)\cos(\lambda^{1/2} N) - 2 a' a''}{2\sin(\lambda^{1/2}N)} \right)\, .
\end{align}
The poles of $S(N)$ are 
\begin{align}
N=N_m\equiv\frac{\pi m}{\lambda^{1/2}}\, 
\end{align}  
for integer $m$.
The residue of the $m=0$ pole is 
\begin{align}
R_{m=0} =  \frac{1}{4}\left( (T''-T')^2 +  (a''-a')^2 \right)\, ,
\end{align}
while the residues of the $m\ne 0$ poles are
\begin{align}
R_{m\ne 0} =   \frac{1}{4}\left(a' - (-1)^m a''\right)^2 \, .
\end{align}
A Green's function satisfying 
\begin{align}\label{hdelt}
\hat H G = \delta(a'-a'')\delta('T'-T'')
\end{align}
 is obtained from contour integrations bounded at the $N=0$ pole.   
The sum over Lefschetz thimbles  may include endpoints at the other finite poles, but only in pairs such that  
there are no additional delta function sources of the form $\delta(R_{m\ne 0})$ on the right hand side of \eqref{hdelt}.  The curves $R_{m\ne 0}=0$ are ghost sources, shown as horizontal lines in figure \ref{cuspchannel1}, to which the $A_3$ caustics are bound. For reasons discussed in section \ref{polesandcaustics}, the location of these caustics along a ghost source is determined by the vanishing of the  
$\mathcal{O}(N-N_m)$ term in the Laurent expansion of the lapse action \eqref{sbarchan} about the $m$'th pole; 
\begin{align}
g_1(a,T) = (T''-T')^2 - \left(\frac{\pi m}{\sqrt\lambda}\right)^2  \left(1 - \lambda\frac{1}{2}(a''^2 + a'^2) \right) =0\, .
\end{align}
This localization can be seen in figure \ref{cuspchannel1}.

Because the path integral here is Euclidean,  the caustics of the Wheeler-DeWitt wave function in the model described here are analytic continuations of the $A_3$ diffraction catastrophe.  For a Lorentzian path integral giving the canonical diffraction catastrophes, the codimension $d\ge 2$ caustics have an important physical property besides large field amplitude;  strong entanglement with respect to partitions of their unfolding degrees of freedom.  The entanglement entropy and its relation to finite $N$ singularities is the topic of the subsequent sections. 



\section{Caustics, dislocations and entanglement}\label{disloc}

\begin{figure}[h!] 	
	\begin{center}
		\begin{subfigure}{.5\textwidth}		
			\includegraphics[width= 210pt]{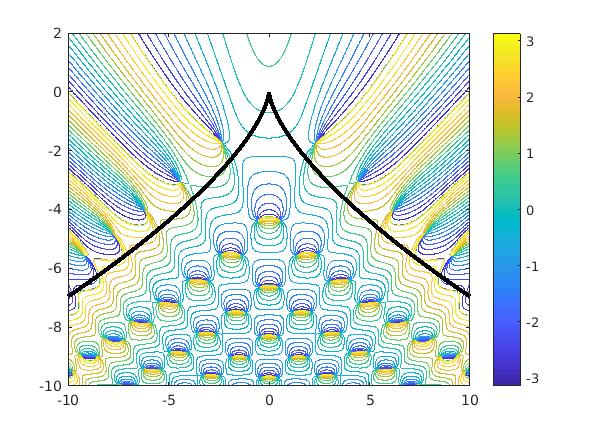}		
		\end{subfigure}\hfill\hskip-50pt
		\begin{subfigure}{.5\textwidth}
			\includegraphics[width= 210pt]{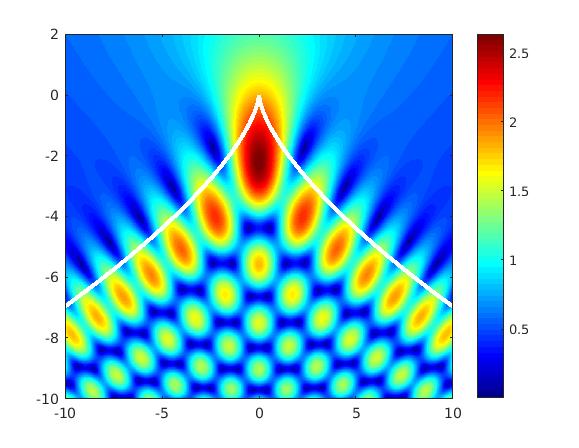}
		\end{subfigure} 
		\caption{ (a) Constant phase contours of the Pearcey function, shown for integer multiples of $\pi/8$, with the cusp caustic curve superimposed.  Wavefront dislocations can be seen as the points at  which  contours meet, around which the phase varies by $2\pi$. The cusp point lies at the origin. \\ \noindent (b) Amplitude of the Pearcey function, with cusp caustic curve superimposed. The defects are the darkest points, at which the wave functions vanish.}\label{fig:Pearcey}
	\end{center}
\end{figure}

The distinguishing observable feature of caustics is usually taken to be the large amplitude of the wave function in their vicinity, which becomes infinite in a classical or short wavelength limit.  However caustics with codimension greater than one have an additional important distinction; they are accompanied by an aggregation of wavefront dislocations, or vortices, at which the wave function vanishes \cite{BerrySing} and around which $\oint d\ln(\Psi) =2\pi n$.   
The dislocations of the $A_3$ diffraction catastrophe can be seen in figure \ref{fig:Pearcey}(a), which shows both the caustic curve and the constant phase contours of the Pearcey function, 
\begin{align}
\Psi(q_1,q_2) \equiv \int d\lambda e^{i(\lambda^4 + q_1\lambda^2 + q_2 \lambda)}\, .
\end{align}
The amplitude is shown together with the caustic curve in figure \ref{fig:Pearcey}(b).  
To the two well known physical characteristics of codimension $d\ge 2$ caustics, large amplitude and dislocations, we will add a third  which arises from them: strong entanglement with respect to partitions of the unfolding directions. 

The entanglement entropy with respect to a partition into the unfolding directions $q_1$ and $q_2$ can be written as a non-local functional of $\Psi$, dependent on $\Phi_{12}\equiv\partial_1\partial_2 {\rm ln}\Psi$ and $|\Psi|$, where the absence of entanglement would imply $\Phi_{12} =0$ everywhere. 
Approaching a dislocation, $\Phi_{12} \rightarrow\infty$ and $|\Psi|\rightarrow 0$,
whereas $\Phi_{12}$ is comparatively small where the wave function is large.  
Non-local terms which contribute substantially to the entropy, in the integral representation to be described below, involve a combination of the dislocation and large field regions. For a wave function containing cusp caustics, the dominant contributions are from combinations of these regions near the cusp points.  



\begin{figure}[h!] 	
	\begin{center}
			
			\includegraphics[width= 210pt]{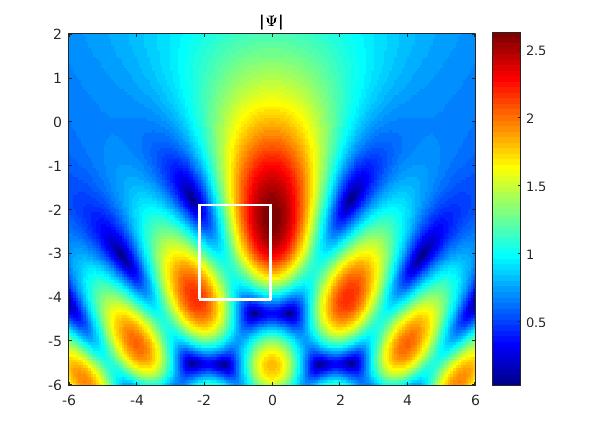}		
			\includegraphics[width =210pt]{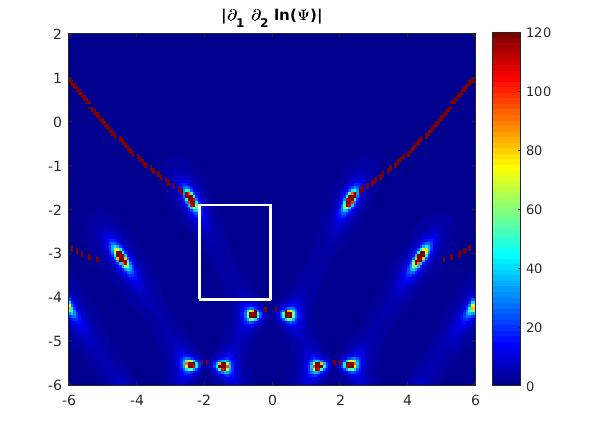}
		
		\caption{The rectangle superimposed on the Pearcey function is that for which $\Upsilon(\Box)$, which defines the measure associated with the entanglement entropy \eqref{boxentcontrib}, is maximal.  
		The figure on the left is the absolute value of the Pearcey function $|\Psi|$, while that on the right is $|\partial_1\partial_2 \ln(\Psi)|$.  In the latter case, the curves emanating from the dislocations are an artifact of the branch choice. Note that the corners of the rectangle connect regions of large field and regions in the neighborhood of dislocations.  
		There is also a another maximal rectangle, related by reflection symmetry. }\label{fig.maxent}
	\end{center}
\end{figure}

The R\'enyi entanglement entropy is defined by
\begin{align}
{\cal R}_{\alpha} \equiv \frac{1}{1-\alpha}\ln \left( {\rm tr}\, \hat \rho^\alpha \right)
\end{align}
where $\hat \rho$ is the density matrix obtained by a partial trace over one of the partitions of the Hilbert space.  Tracing over $q_2$ gives
\begin{align}
\hat \rho_{q_1,q_1'} = \frac{1}{{\cal M}^2}\int dq_2 \Psi(q_1,q_2)\Psi^{*}(q_1',q_2)\, ,
\end{align}
where ${\cal M}$ is a normalization factor. 
The diffraction catastrophes are not normalizable, but this fact will be temporarily ignored. 
Tracing over $q_1$ instead gives the same result for the R\'enyi entropy.
The simplest case to consider is ${\cal R}_2= - \ln({\rm tr}\hat \rho^2)$.
Writing the wave function as $\Psi =  A\exp(i\Theta)$ where $A\equiv |\Psi|$,
\begin{align}\label{R2D2}
{\rm tr}(\hat \rho^2) &= \frac{1}{{\cal M}^4}\int dq_1 dq_2 dq_1' dq_2' A(q_1,q_2)A(q_1,q_2')A(q_1',q_2')A(q_1',q_2)\nonumber \\
 &{} \hskip 40pt \exp\left(i \left(\Theta(q_1,q_2)-\Theta(q_1,q_2')+\Theta(q_1',q_2')-\Theta(q_1',q_2)\right)\right)\, . \end{align}
It is convenient to write this in a short hand fashion, as an integral over rectangles in the $q_{1,2}$ plane,
\begin{align}\label{boxexp}
{\rm tr}(\hat \rho^2) =   \frac{1}{{\cal M}^4} \int d\Box  A_4(\Box) \, \exp\left(i\int_\Box dq_1 dq_2\, \partial_1\partial_2 \Theta \right)\, ,
\end{align}
where 
\begin{align}
d&\Box \equiv dq_1 dq_2 dq_1' dq_2' \nonumber \\
A&_4(\Box) \equiv A(q_1,q_2)A(q_1,q_2')A(q_1',q_2')A(q_1',q_2)\, .
\end{align}
The expressions occuring for the higher R\'enyi entropies are similar, with  rectangles replaced by right angle $2n-$gons, which may be non-convex and self intersecting.  Writing ${\rm tr}\left(\hat\rho^2\right) = 1-\chi$, the entanglement is characterized by how much $\chi$ differs from 0.  The quantity $\chi$ is obtained from a measure over the set of rectangles, $\chi = \int d\mu(\Box)$ where  
\begin{align}\label{boxentcontrib}
d\mu(\Box) &= d\Box\frac{1}{{\cal M}^4}\Upsilon(\Box)\\
\Upsilon&(\Box) \equiv A_4(\Box) \, \left[\exp\left(\int_\Box dq_1 dq_2\, \partial_1\partial_2 \ln(A) \right)
- \exp\left(i\int_\Box dq_1 dq_2\, \partial_1\partial_2 \Theta \right) \right]\, . \nonumber 
\end{align}
In writing the above expression we have used the fact that ${\cal M}^2 = \int dq_1 dq_2 A(q_1,q_2)^2$ implies 
\begin{align}\label{n4}
{\cal M}^4 =  \int d\Box  A_4(\Box) \, \exp\left(\int_\Box dq_1 dq_2\, \partial_1\partial_2 \ln(A) \right)\, .
\end{align}
Note that there are four ways of choosing opposing corners $(q_1,q_2)$ and $(q'_1,q'_2)$ defining a given rectangle;  summing over these shows that $\Upsilon(\Box)$ is real and positive. 
Diffraction catastrophes are  not normalizable, so that the entanglement entropy is ill defined without regularization. Nevertheless, ignoring the question of regularization, the term $\Upsilon(\Box)$ in the definition of the measure \eqref{boxentcontrib} is well defined; 
we shall seek the regions in which it is maximal.
The maximum  can be found by numerical computation for all rectangles within a discretized domain. 
The maximal rectangle, shown in figure \ref{fig.maxent}, is 
near the cusp point, with corners containing both large amplitude points and points near dislocations.
In this sense the twin features of a codimension $d\ge 2$ caustic, large fields and dislocations, give substantial contributions to the entanglement for a partition defined with respect to the unfolding directions.  

   
\begin{figure}[h!] 	
	\begin{center}
		\begin{subfigure}{.5\textwidth}		
			\includegraphics[width= 200pt]{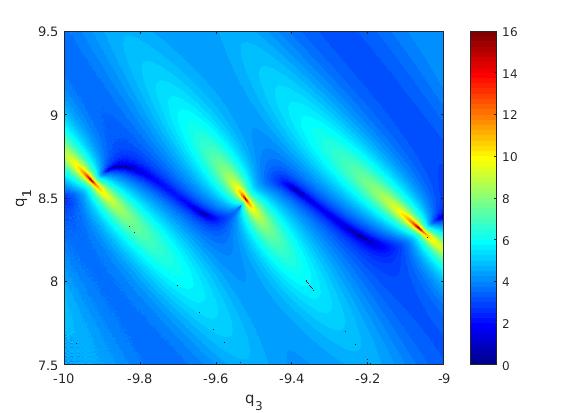}	
			\caption{}	
		\end{subfigure} \hfill\hskip-10pt
		\begin{subfigure}{.5\textwidth}
			\includegraphics[width= 200pt]{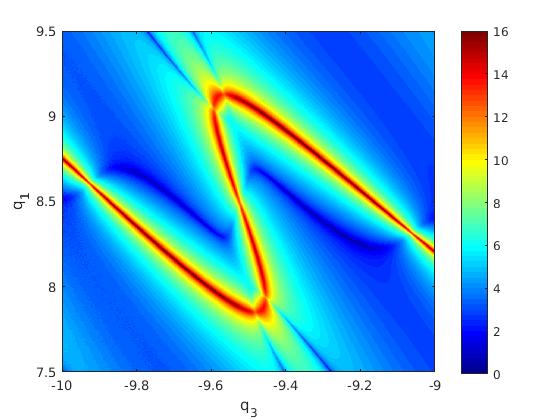}
			\caption{}
		\end{subfigure}
		\caption{ (a) Tripartite entanglement requires non-zero $\ln |\partial_1\partial_2\partial_3 (\ln\Psi)|$, shown here for the swallowtail diffraction catastrophe  \eqref{swa}, evaluated along a section $q_2=0.3$.  
			This quantity diverges approaching  points at which the section is intersected by dislocation strings transverse to the page.   (b) $\ln |\partial_1\partial_2\partial_3(\ln\Psi)|$ along the section $q_2=0$.  The wavy curve seen here corresponds to  a dislocation lying entirely within the $q_2=0$ plane.  There are string junctions at which this dislocation intersects those seen in (a). }\label{tripart}	
	\end{center}
\end{figure}

Thus far we have considered entanglement for a caustic unfolding in two dimensions.  The higher dimensional case is also interesting. 
Since dislocations are codimension two,
they form webs of strings for a caustic unfolding in three dimensions. The dislocations have been studied in detail for the swallowtail caustic or $A_4$ catastrophe \cite{NyeSwallowtail}, with diffraction catastrophe given by   
\begin{align}\label{swa}
\Psi= \int d\lambda \exp\left( i \left( \frac{1}{5}\lambda^5 + \frac{1}{3}q_3\lambda^3 + \frac{1}{2}q_2\lambda^2 + q_1\right) \right)\, .
\end{align}
In this case there is also tripartite entanglement related to non-vanishing $\Gamma_{123} \equiv \partial_1\partial_2\partial_3 (\ln\Psi)$.  
Figure \ref{tripart} shows the behavior of  $\Gamma_{123}$  on some sections of fixed $q_2$. Although it is possible for  $\Gamma_{123}$ to vanish along dislocation strings, depending on the choice of partition and the curvature of the strings, it necessarily diverges at a string junction.


\section{Clocks and caustics}\label{clocstics}

Strong entanglement among the unfolding degrees of freedom at a codimension $d\ge 2$ caustic is an intriguing phenomenon from the point of view of the Wheeler-deWitt wave function when one of the unfolding directions can be interpreted as a physical clock.  
In the ADM formalism, 
the Hamiltonian constraint means that wave function is static; the notion of time evolution is only defined with respect to entanglement\cite{PageWooters,Wooters} between a physical clock $T$ and the rest of the degrees of freedom $R$, enforced by $\hat H\Psi =0$.  
In the simplest case, the Hamiltonian breaks up into a sum of commuting parts, $\hat H= \hat \pi_T + \hat H_R$, where $[\pi_T,T] = i$.  
The entangled state satisfying the Hamiltonian constraint is 
\begin{align}
|\Psi\rangle = \int dT\,\, |T\rangle \otimes |\Psi(T)\rangle_R
\end{align}
%
where
\begin{align}
(i\partial_T - \hat H_R)|\Psi(T)\rangle_R = 0\, .
\end{align}
Note that if the clock Hamiltonian $\hat H_{clock}$ and $\hat H_R$ commute but $\hat H_{clock} \sim \pi_T^2$, as in the model of section \ref{examples},  the Hamiltonian constraint $(\hat H_{clock} + \hat H_R)|\Psi\rangle = 0$ requires 
\begin{align}
\left(i \partial_T - \sqrt{\hat H_R}\right)\left(i\partial_T + \sqrt{\hat H_R}\right)|\Psi(T)\rangle_R = 0\, .
\end{align}
A Shr\"odinger equation still arises
provided one only considers solutions propagating forward in the clock time
with respect to the Hamiltonian $\sqrt{\hat H_R}$.  
The problem of  defining ``good clocks'' and extracting the Schr\"oedinger equation from quantum gravity \cite{Isham} will not be considered here.  Our contribution to the relation between entanglement and time concerns the case in which a clock is also an unfolding degree of freedom of a caustic.

  
Suppose that the Wheeler-DeWitt wave function has an $A_3$ caustic with unfolding directions $T$ and $Q$, and that there may be other degrees of freedom as well; $Q\in R$. 
For a partition between $T$ and $Q$, the measure associated with the integral representation of the R\'enyi entanglement entropy has largest contribution from a neighborhood of the cusp due to a conspiracy between regions of large field and dislocations.
The latter lie solely in the $Q,T$ plane. Since dislocations are codimension two, the same mechanism of entanglement can not occur for a partition between $Q$ and other variables in $R$, analogous to the monogamy of entanglement. 
In becoming strongly entangled with the clock, one expects $Q$ to be disentangled with other degrees of freedom in $R$. Furthermore,  the focal properties of the caustic collapse the wave function with respect to $Q$. Thus the caustic bears some resemblance to a quantum measurement. 
The instant $T$ of ``measurement'' maps to a pole of the lapse action while the value of the collapsed $Q$
is that at which the residue of the pole vanishes. 

The above comparison with a quantum measurement assumes a Lorentzian path integral.  The example of section \ref{examples} contains  $A_3$ caustics in the plane spanned by the scale factor and a scalar field behaving as a clock, but the Euclidean path integral yields an analytic continuation of the $A_3$ diffraction catastrophe.  Dislocations are absent in the Euclidean case and the above discussion of entanglement does not apply.


\section{Lefschetz thimbles and entanglement}\label{ThimbEntangle} 

The R\'enyi entanglement entropy, defined with respect to partitions of unfolding directions of a caustic, can be expressed in terms of integrals over multiple lapse variables,
\begin{align} {\cal R}_n = \frac{1}{1-n}{\rm ln}\left(\int dN_1\cdots dN_{2n}e^{i\Gamma(N_1\cdots N_{2n})}\right)\, .
\end{align}
We show below that $\Gamma$, referred to as the R\'enyi action, has a singularity structure different from that of the lapse action ${\mathbb S}$.
Although the essential singularities of $\exp(i\Gamma)$ at infinity are inherited from those of $\exp(i{\mathbb S})$,
the finite poles of ${\mathbb S}$ are absent in $\Gamma$,
replaced with non-essential singularities of $\exp(i\Gamma)$  at $N_i=N_j$ for cyclic pairings of the indices. 
The Lefschetz thimbles associated with $\exp(i\Gamma)$ evade the non-essential singularities in the interior and are bounded by the essential singularities at infinity, such that the relative homology classes they represent are a higher dimensional variant of a link.


Consider the example of an $A_3$ diffraction catastrophe given in section \ref{Lefschetz},   
\begin{align}\label{cuspwavefunc}
\Psi(q_1,q_2) &= \int dN e^{i{\mathbb S}(N)} \nonumber\\
&= \int dN \sqrt{\frac{i\mu}{ 4\pi N(N-\mu)}} \exp\left[ i\left( \frac{1}{4(N-\mu)}q_1^2 + \frac{1}{4N}q_2^2 + \alpha N\right)\right]\, .
\end{align}
The R\'enyi entanglement entropy will be defined with respect to a density matrix 
obtained by a partial trace over $q_2$, 
 \begin{align}\label{partrace}
 \rho_{q_1,q'_1} &= \int dq_2 \Psi(q_1,q_2)\Psi^* (q'_1,q_2) \, ,
 \end{align}
 where for the moment we neglect the non-normalizability of \eqref{cuspwavefunc}.
Expressing the R\'enyi entropy ${\cal R}_2$ using \eqref{cuspwavefunc} and \eqref{partrace} and carrying out the integrations over $q$ yields a remarkably simple result,  in which the finite $N$ essential singularities $N=0$ and $N=\mu$ are absent;
\begin{align}\label{expgam}
 {\cal R}_2
 \equiv & -{\rm ln}\,{\rm tr}(\hat \rho^2)=
 -{\rm ln}\left( \int dN_1  dN_2   dN_3  dN_4\,  e^{i\Gamma(N_1,N_2,N_3, N_4)}\right)\, , \nonumber\\
e^{i\Gamma}&=\frac{1}{{\cal M}^4}
\left[  
(N_2-N_1)(N_3-N_2)(N_4-N_3)(N_4-N_1)
\right]^{-1/2} 
e^{i\alpha(N_1-N_2+N_3- N_4)}\, ,
\end{align}
where ${\cal M}$ is a normalization factor.
Similar expressions apply to the higher order R\'enyi entropies 
${\cal R}_{n>2}$.  
In addition to its derivation from a Lorentzian path integral, the result \eqref{expgam} depends critically on the parameters $\mu$ and $\alpha$ being real, so that $\Psi$ in the neighborhood of the cusp caustic is described by the standard diffraction catastrophe rather than some analytic continuation. 

The finite poles of ${\mathbb S}$ are absent in $\Gamma$  because their residues are dependent on $q_i$, over which one integrates to obtain $\Gamma$.  To illustrate, the integration contains terms of the form  
\begin{align}\label{divterm}
e^{i\Gamma} = \int dq_1\,\exp\left(i \frac{q_1^2}{4} \left( \frac{1}{N_i-\mu} -  \frac{1}{N_j-\mu} \right) \right)\cdots \, ,
\end{align}
which diverges when $N_i=N_j$.  This is the mechanism by which the
finite $N$ essential singularities of $\exp(i\mathbb S)$ become non-essential $N_i=N_j$ singularities of $\exp(i\Gamma)$.  Lefschetz thimbles avoid these rather than ending on them.  
The boundaries of the Lefschetz thimbles are the essential singularities of $\exp(i\Gamma)$ at infinity,  which are inherited trivially from those of $\exp(i{\mathbb S})$ due to their lack of dependence on $q_i$.  

As noted in section \ref{disloc}, the diffraction catastrophes are not normalizable, so the R\'enyi entanglement entropy is ill defined in the absence of regularization.
The integral over $N_i$ in \eqref{expgam} is divergent due to invariance of the integrand under a constant shift of all the $N_i$. 
Formally one can write
\begin{align}\label{Fterm}
{\cal R}_2=-\ln&\left( 
\frac{{\rm tr}\hat\rho^2} {({\rm tr}\hat\rho)^2} \right) = -{\rm ln}\left(\frac{E}{F^2}\right) \nonumber\\
E \equiv \int&  dN_1 d N_2 dN_3 d N_4
\left[  
( N_2-N_1)(N_3- N_ 2)( N_4-N_3)( N_4-N_1)
\right]^{-1/2} \nonumber\\ 
& e^{i\alpha(N_1- N_2+N_3- N_4)} \nonumber\\
F  \equiv \int &   dN_1 d N_2 (N_1- N_2)^{-1}e^{i\alpha(N_1- N_2)}\, .
\end{align} 
Thus a regularization  
$\int dN \rightarrow 1/\epsilon$ implies ${\cal R}_2 = -{\rm ln}(\epsilon) + {\rm finite}$. In general, ${\rm tr}(\hat\rho^n)/({\rm tr}\hat\rho)^n \sim \epsilon^{-1+n}$ so that
\begin{align}
{\cal R}_n = \frac{1}{1-n}
\ln\left( 
\frac{{\rm tr}\hat\rho^n} {({\rm tr}\hat\rho)^n} \right)
 = -{\rm ln}(\epsilon) + {\rm finite}\, ,
\end{align}
where the finite part is determined by the integral over the subspace of $\{N_i\}$ on which the Lefschetz thimbles are defined.
Note that the gradient flow method to obtain the Lefschetz thimbles from any initial integration cycle, described in \cite{TurokRadio}, can be carried out even when the integral is not convergent.



\begin{figure}[!h]
	\center{
		\includegraphics[width= 150pt]{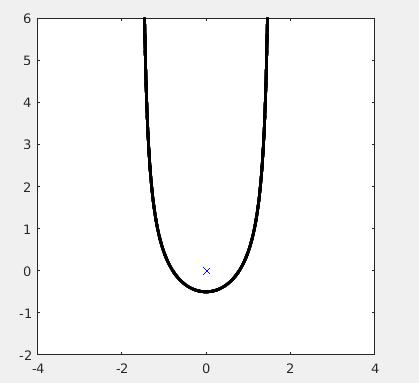}
		\caption{A Lefschetz thimble associated with \eqref{simpL} in the complex $N_- \equiv N_1-N_2$ plane, obtained by gradient flow from the initial surface \eqref{initsurf}.}
		\label{fig:LLink}
	}
\end{figure}

\begin{figure}[!h]
	\center{
		\includegraphics[width= 150pt]{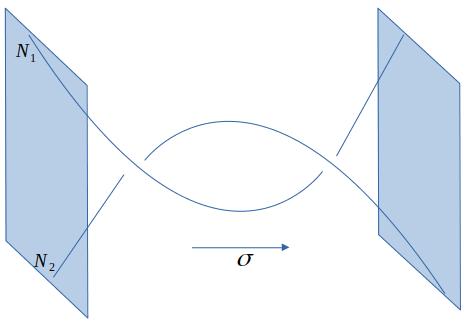}
		\caption{Curves $N_i(\sigma)$  associated with the Lefschetz thimble obtained by gradient flow with respect to $Re(\Gamma)$, starting from the initial surface \ref{initsurf}.  This becomes a Hopf link upon identifying the points at $\Im(N_1) = \infty$ and also the points at $\Im(N_2)=-\infty$}
		\label{fig:braid}
	}
\end{figure}

The term $F$ in \eqref{Fterm} provides a simplified example of the geometry of Lefschetz thimbles associated with the R\'enyi action.  Consider Lefschetz thimbles defined with respect to $\exp(i\Gamma)$ where
\begin{align}\label{simpL}
\Gamma= -(1/2)\log(N_1-N_2) +i\alpha(N_1-N_2)\, .
\end{align}
Although $\int dN_1 dN_2 \exp(i\Gamma)$ is divergent for the reasons discussed above, Lefschetz thimbles are defined in the complex $N_-$ plane with  $N_- \equiv N_1-N_2$.  One approach to finding Lefschetz thimbles is the gradient flow method described in \cite{TurokRadio}.  To this end, consider an initial surface 
\begin{align}\label{initsurf}
N_-(\sigma,\rho) &\equiv N_1(\sigma,\rho)-N_2(\sigma,\rho) =\sigma-i\delta \nonumber \\ 
N_+(\sigma,\rho) &\equiv   N _1(\sigma,\rho)+N_2(\sigma,\rho)  =\rho\, ,
\end{align}
for a small positive number $\delta$ and real parameters $\sigma$ and $\rho$.
The gradient flow of ${\rm Re}(\Gamma)$ depends only on $N_-$, yielding the curve shown in figure \ref{fig:LLink}.  The new surface can be written as
\begin{align}\label{braideqn}
2N_1(\sigma,\rho) &= \rho + N_-(\sigma) \nonumber\\
2N_2(\sigma,\rho) &= \rho - N_-(\sigma) \, ,
\end{align}
where $N_-(\sigma)$ is a parameterization of the curve shown in \ref{fig:LLink}.
The Lefschetz thimble is bounded at $N_1=+i\infty$ and $N_2= -i\infty$, encircling the $N_1=N_2$ singularity in the interior.  For any constant $\rho$, which corresponds to the value of $N_+$ on which the integrand does not depend, \eqref{braideqn} represents a Hopf link shown in figure \ref{fig:braid}, in which the points $\Im(N_1)=\infty$ are identified, as well as the points $\Im(N_2)=-\infty$. 

The above example is simplified for the sake of illustration. More generally, the Lefschetz thimbles\footnote{It is not obvious how to map the relative homology classes for integration cycles in the case of the lapse action ${\mathbb S}(N)$ to those associated with the R\'enyi action $\Gamma_n(N_1\cdots N_{2n})$, and no attempt will be made here.} associated with ${\rm tr}(\hat\rho^n)$
have the form
\begin{align}
N_{i}&(\rho,\sigma_1\cdots\sigma_{2n-1}) =\frac{1}{2n} \rho + f_i(\sigma_1 \cdots \sigma_{2n-1}) \\
&\sum_{i=1}^{2n}f_i(\sigma_1 \cdots \sigma_{2n-1}) = 0
\end{align}
for $i = 1\cdots 2n$, where $\rho,\sigma_{1\cdots 2n-1}$ are real.  
The complex functions $f_i(\sigma_1\cdots \sigma_{2n-1})$ define
a $2n-1$ dimensional manifold embedded in $2n+1$ dimensions which is a higher dimensional variant of a link. 
This is a variant not just in the sense of dimensionality, but also because the configuration space is somewhat different from that of the braid group.  The latter 
is given by
\begin{align}
\{N_1,\cdots,N_m\}:N_i \in {\mathbb C}, N_i\ne N_j, \,{\rm for}\,\, i\ne j \, .
\end{align}
However the form of the R\'enyi action, such as \eqref{expgam}, implies that the $N_i$ lie on a multiply sheeted cover of ${\mathbb C}^m$, on which only cyclic pairings are removed,
\begin{align}
N_i \ne N_j,\, {\rm for}\, j = i\pm 1\, {\rm mod}\, m
\end{align}   

The R\'enyi action \eqref{expgam}  is specific to the particular representation of a cusp caustic \eqref{cuspwavefunc}. For a general wave function,  there may be a network of caustics, of which this result is a zoomed in view.  A natural question is to what extent the singularities 
at $N_i = N_j$ persist in the more general case.   The answer depends on the partitioning of the Hilbert space,  but there are arguments that these singularities are present so long as one of the partial traces includes an integration along a degree of freedom $\xi$ transverse to a ghost source.   The integrals giving the R\'enyi action then contain terms similar to \eqref{divterm} of the form 
\begin{align}
e^{i\Gamma}=
 \int d\xi e^{\frac{i}{4}  \left(     \frac{\xi^2}{N_i-\mu} - \frac{\xi^2}{N_j-\mu} 
				        \right)    
				   }\cdots 
\end{align} 
so that the integral over $\xi$ removes the essential singularities at $N = \mu$ and diverges when $N_i =N_j$.  
Consider the example of a Lorentzian variant of \eqref{sbarchan}, for which there are an infinite number of cusp caustics and the lapse action contains an infinite number of poles in the complex $N$ plane.  Partitioning the Hilbert space into the scale factor $a$ and the clock variable $T$,  
 \begin{align} \label{gammNint}
{\rm tr}(\hat \rho^2) &= \int dN_1 d N_2 dN_3 d N_4 e^{i\Gamma(N_1, N_2, N_3, N_4)}\nonumber\\ 
&= \int  d a_1 dT_1 da_2 d T_2 \Psi(T_1,a_1)\Psi^*(T_2,a_1) \Psi(T_2,a_2)\Psi^*(T_1,a_2) 
\end{align}
where $a$ is equivalent to $\xi$, the degree of freedom transverse to the ghost source and
\begin{align} \label{psii}
\Psi(a,T) = \int dN  \frac{1}{(N \sin(\lambda^{1/2}N))^{1/2}} \exp\left[   
	-i\left(\frac{T^2}{4N} +\frac{1}{2} N + \lambda^{1/2}\frac{ a^2\cos(\lambda^{1/2} N) }{4\sin(\lambda^{1/2}N)} \right)\right]\, .
\end{align}
To arrive at \eqref{psii}, the boundary conditions in \eqref{Nsup}  have been chosen to be $a(0) = 0\, , T(0)=0,\, a(1)=a,\,T(1)=T$.   
Using \eqref{psii} and carrying out the integrals over $a$ and $T$ in \eqref{gammNint} gives,   
\begin{align}
e^{i\Gamma_2} = &\left[ ( N_4 - N_1)( N_2 - N_3) \sin(\lambda^{1/2}( N_4 - N_3)) \sin(\lambda^{1/2}( N_2- N_1)) \right]^{1/2} \nonumber \\
&e^{-\frac{i}{2}(N_1- N_2 + N_3 - N_4)}\, .
\end{align}
The associated Lefschetz thimbles are bounded by essential singularities of $\exp(i\Gamma)$ at infinity 
and entwined in the interior by circumventing  cyclic non-essential singularities at $N_i =N_j$ or  
$N_i = N_j \mod\sqrt{\lambda}\pi$.



\section{Summary and Discussion}

The wave function in diffeomorphism invariant quantum theories can be expressed as a complex integral $\int dN \exp(i {\mathbb S})$, where $N$ is a gauged fixed Lagrange multiplier enforcing the vanishing Hamiltonian constraint, known as the lapse in the context of quantum gravity.  We have shown that  finite $N$ singularities of the lapse action  ${\mathbb S}(N)$ are related to diffraction catastrophes with codimension greater than one.  Finite $N$ poles of ${\mathbb S}$ are associated with $A_{n\ge 3}$ caustics bound to the curves along which the residues of the poles vanish.  The integration contour includes oppositely oriented pairs of Lefschetz thimbles terminating at these poles, yielding the sum of a source and a sink at the locus of vanishing residue, referred to here as a ghost source. A variant of a minisuperspace solution of the Wheeler-DeWitt equation originally described in \cite{HalliwellMyers} was given to illustrate the relation between poles of the lapse action and $A_3$ caustics.  
 
Although caustics are generally understood as regions in which the wave function is very large,   caustics of codimension $d \ge 2$ 
are also accompanied by networks of wave front dislocations.
The R\'enyi entanglement entropy, defined 
with respect to partitions among unfolding directions, 
has an integral representation in which there are substantial contributions from a non-local collusion between regions of large field and regions of rapid phase change near a dislocation.  Caustics of codimension $d\ge 3$ also exhibit substantial contributions to multi-partite entanglement, due to a collusion between  regions of large field and regions where dislocations join or intersect, such as at string junctions if $d=3$. 

    

In the context of the Wheeler-DeWitt wave function, the association of a physical clock with an unfolding degree of freedom of a codimension $d\ge 2$ caustic is particularly interesting. At an $A_3$ caustic for which a clock $T$ parameterizes a ghost source, another unfolding degree of freedom $Q$ transverse to the ghost source behaves in many ways like a quantum mechanical observable being measured.  The singularity of the lapse action, to which the caustic owes its existence, corresponds to the collapse of the wavefunction with respect to $Q$. There is a one to one map between the time of ``measurement'' and the the singular values of the lapse.
In the neighborhood of the caustic, $Q$ becomes highly entangled with $T$ and disentangled with other degrees of freedom,  where the strength of entanglement is defined with respect to a non-local measure which couples regions of large field and dislocations.

The R\'enyi entanglement entropy can also be written as an integral over lapse variables, ${\cal R}_n = \frac{1}{1-n}{\rm ln} \int dN_1 \cdots dN_{2n} \exp(i\Gamma_n(N_1 \cdots N_{2n}))$. 
Although the essential singularities of $\exp(i\Gamma)$ at infinity are inherited from those of $\exp(i{\mathbb S})$,
the finite $N$ essential singularities of $\exp(i\mathbb S)$ are absent in $\exp(i\Gamma)$, having been replaced by poles and branch points occurring when $N_i=N_j$ for cyclic pairings of $i$ and $j$.
The Lefschetz thimbles terminate at essential singularities at infinity while evading the non-essential singularities at $N_i = N_j$, such that the relative homology classes they represent are higher dimensional variants of links.
The $N_i = N_j$ singularities of $\exp(i\Gamma)$ arise from poles in ${\mathbb S}(N)$ upon integrating over a degree of freedom which parameterizes the residue, i.e. the unfolding direction of a caustic which is transverse to a ghost source.  In other words, the singularities of the lapse action responsible for the presence of  codimension $d\ge 2$ caustics are also behind the higher dimensional link geometry of Lefschetz thimbles in the lapse integral representation of the R\'enyi entropy.

%

Examples of a wave function of the universe containing $D_n$ or $E_n$ catastrophes
are currently unknown. 
These have corank $>1$, rendering a map from their potential functions to a function of a single lapse variable unclear.  
However the ADM formalism of quantum gravity \cite{ADM} contains multiple Lagrange multipliers associated with  diffeomorphism invariance, the lapse $N$ and the shift variables $\beta^i$. One can define an action ${\mathbb S}(N,\beta^i)$ generalizing the lapse action, for which it is natural to conjecture the existence of singularities associated with higher corank caustics. While singularities of the lapse action ${\mathbb S(N)}$ imply the collapse of the wave function with respect to global degrees of freedom,  singularities of ${\mathbb S}(N,\beta^i)$ could correspond to local collapse, due to 
unfolding directions interpretable as rulers as well as a clock. 

\section{Acknowledgments}

I would like to thank Chuck Spofford and Katherine Woolfe  for discussions and collaboration on related work.

\end{document}